\documentclass[11pt]{article}

\usepackage{amsmath,amsfonts,latexsym,fullpage,graphicx,vmargin,natbib}
\usepackage{amssymb, mathrsfs}
\usepackage{pstricks}
\usepackage{subcaption}
\usepackage{amsmath}
\usepackage{multirow}
\usepackage{comment}
\usepackage[inline]{enumitem}
\usepackage{hyperref}
\usepackage[normalem]{ulem}
\usepackage{placeins}

\usepackage{enumerate}
\usepackage{pstricks}
\usepackage{subcaption}
\usepackage{multirow}
\usepackage{comment}
\usepackage[inline]{enumitem}
\usepackage{amsfonts}
\newcommand{\Var}{\mathrm{Var}}

\newcommand{\iid}{\stackrel{\mbox{\scriptsize iid}}{\sim}}
\newcommand{\ind}{\stackrel{\mbox{\scriptsize ind}}{\sim}}
\newcommand{\bm}[1]{\mbox{\boldmath{$#1$}}}

\newcommand{\calN}{\mathcal{N}}

\newcommand{\Ptilde}{\widetilde{P}}
\newcommand{\Gtilde}{\widetilde{G}}

\newcommand{\Rea}{{\mathbb{R}}}
\newcommand{\Law}{\mathcal L}

\newcommand{\cov}{\text{cov}}
\newcommand{\Pp}{\mathbb{P}}

\newcommand{\Y}{\mathbb{Y}}
\newcommand{\E}{\mathbb{E}}
\newcommand{\R}{\mathbb{R}}
 
\usepackage{mathtools}
\newcommand{\bdot}{\boldsymbol{\cdot}}
\newtheorem{theorem}{Theorem}[section]

\newtheorem{prop}[theorem]{Proposition}

\newcommand{\virgolette}[1]{``#1''}

\renewcommand{\mid}{\,|\,}

\newcommand{\qed}{\hfill \ensuremath{\Box}}

\makeatletter

\begin{document}

\title{\bf The semi-hierarchical Dirichlet Process and its application to clustering homogeneous distributions}
\author{
  Mario Beraha\thanks{Department of Mathematics, Politecnico di Milano}
  \thanks{Department of Computer Science, Universit\`a di Bologna},
  Alessandra Guglielmi\footnotemark[1],
  Fernando A. Quintana\thanks{Department of Statistics, Pontificia Universidad Cat\'olica de Chile}\thanks{ANID - Millennium Science Initiative Program - Millennium Nucleus
  Center for the Discovery of Structures in Complex Data}
}
\date{\today}

\maketitle

\begin{abstract}
    Assessing homogeneity of distributions is an old problem that has received
    considerable attention, especially in the nonparametric Bayesian
    literature. To this effect, we propose the semi-hierarchical Dirichlet
    process, a novel hierarchical prior that extends the hierarchical
    Dirichlet process of~\cite{Teh_etal_HDP_2006} and that avoids the
    degeneracy issues of nested processes recently described
    by~\cite{Cam_etal_2018latent}. We go beyond the simple yes/no answer to
    the homogeneity question and embed the proposed prior in a random
    partition model; this procedure allows us to give a more comprehensive
    response to the above question and in fact find groups of populations that
    are internally homogeneous when $I\ge 2$ such populations are considered.
    We study theoretical properties of the semi-hierarchical Dirichlet process
    and of the Bayes factor for the homogeneity test when $I=2$. Extensive
    simulation studies and applications to educational data are also
    discussed.
\end{abstract}

\section{Introduction}\label{sect:intro}

The study and development of random probability measures in models that
take into account the notion of data that are not fully exchangeable has
sparked considerable interest in the Bayesian nonparametric literature. We
consider here the notion of partial exchangeability in the sense of de
Finetti~\citep[see][]{deFinetti1938,diaconis1988recent}, which
straightforwardly generalized the notion of an exchangeable sequence of
random variables to the case of invariance under a restricted class of
permutations. See also~\cite{CAMERLENGHI201718} and references therein. In
particular, our focus is on assessing whether two or more populations (or
groups) of random variables can be considered exchangeable rather than
partially exchangeable, that is whether they arose from a common
population/random distribution or not.

To be mathematically accurate, let us introduce partial exchangeability
for a sequence of random variables. Let $\Y$ denote a complete and
separable metric space (i.e. a Polish space) with corresponding metric
$d$. Let $\mathcal{Y}$ denote the Borel $\sigma$-algebra of $\Y$, and
$\mathbb{P}_{\Y}$ denote the space of all probability measures on
$(\Y,\mathcal{Y})$, with Borel $\sigma$-algebra $\mathcal{P}_{\Y}$. We
will often skip reference to $\sigma$-algebras. A double sequence $(y_{1 1
},y_{1 2 }, y_{1 3}, \ldots, y_{2 1}, y_{2 2 }, y_{2 3}, \ldots)$  of
$\Y$-valued random variables, defined on a probability space
$(\Omega,\mathcal{F},P)$ is called \textit{partially exchangeable} if for
all $n,m\geq 1$ and all permutations $(i(1), \ldots, i(n))$ and $(j(1),
\ldots , j(m))$ of $(1, \ldots , n)$ and $(1, \ldots ,m)$ respectively, we
have
\begin{equation*}
\mathcal{L}(y_{1 1 }, \ldots, y_{1 n}, y_{2 1 }, \ldots, y_{2 m}) =
\mathcal{L}(y_{1 i(1)}, \ldots, y_{1 i(n)}, y_{2 j(1)}, \ldots, y_{2 j(m)}).
\end{equation*}
Partial exchangeability can thus be conceptualized as invariance of the
joint law above under the class of {\em all} permutations acting on the
indices {\em within} each of the samples. Here and from now on, the
distribution of a random element $y$ is denoted by $\mathcal{L}(y)$.

The previous setting can be immediately extended to the case of $I$
different populations or groups. By de Finetti's representation theorem
\citep[see the proof in][]{regazzini1991coherence}, partial
exchangeability for the array of $I$ sequences of random variables $(y_{1
1 }, y_{1 2 }, \ldots, y_{2 1 }, y_{2 2 }, \ldots $, $y_{I 1 }, y_{I 2
},\ldots )$ is equivalent to
\begin{equation*}
P(y_{i j}\in A_{i j},\, j=1,\ldots,N_i,\, i=1,\ldots,
I)=\int_{\mathbb{P}_{\Y}^I} \prod_{i=1}^I \prod_{j=1}^{N_i} p_i(A_{i j})\,
Q(dp_1,\ldots,dp_I),
\end{equation*}
for any $N_1,\ldots,N_I\ge 1$ and Borel sets $\{A_{i j}\}$ for
$j=1,\ldots,N_i$ and $i=1,\ldots,I$. In this case, de Finetti's measure
$Q$ is defined on the $I$-fold product space
$\mathbb{P}_{\Y}^I=\mathbb{P}_{\Y}\times\mathbb{P}_{\Y}\times\cdots\times
\mathbb{P}_{\Y}$, and $(p_1,p_2,\ldots, p_I)\sim Q$. The whole joint
sequence of random variables is exchangeable if and only if $Q$ gives
probability 1 to the measurable set $S = \{(p_1, p_2,\ldots, p_I) \in
\mathbb{P}_{\Y}^I:\, p_1 = p_2=\cdots = p_I\}$.

Hence, partial exchangeability of data from different groups (or related
studies) is a convenient context to analyze departures from
exchangeability. While homogeneity of groups here amounts to full
exchangeability, departures from this case may follow different
directions, including independence of the population distributions $p_1,
p_2,\ldots, p_I$.
However it could be interesting to investigate other types of departures
from exchangeability beyond independence. The main goal here is to build a
prior $Q$ for $(p_1, p_2,\ldots, p_I)$ that is able to capture a wider
range of different behaviors, not only restricting the analysis to
assessing equality or independence among $p_1, p_2,\ldots, p_I$.
In the simplest case of $I=2$, we just compare two
distributions/populations, but we aim here at extending this notion to
$I>2$ groups. In particular, we address the following issue: if the answer
to the question of homogeneity within all these groups is negative, a
natural question immediately arises, namely, can we assess the existence
of homogeneity within certain populations? In other words, we would like
to find clusters of internally homogeneous populations.

Vectors of dependent random distributions appeared first in
\cite{CifReg78}, but it was  in \cite{maceachern1999dependent} where a
large class of dependent Dirichlet processes was introduced, incorporating
dependence on covariates through the atoms and/or weights  of the
stick-breaking representation. Following this line, \cite{DeIorioetal04}
proposed an ANOVA-type dependence for the atoms. These last two papers
have generated an intense stream of research which is not our focus here.
For a review of such constructions, see~\cite{DDPreview}.

Our approach instead constructs a prior that explicitly considers a
departure from exchangeability. Other authors have considered similar
problems. \cite{muller_etal2004} and \cite{Lijoi_etal_CSDA14} constructed
priors for the population distributions by these distributions with the
addition of a common component. See also \cite{hatjispyros2011dependent},
\cite{hatjispyros2016random} and \cite{hatjispyros2018dependent} for
related models with increasing level of generalization. Several references
where the focus is on testing homogeneity across groups of observations
are available.
\cite{ma2011coupling} and \cite{soriano2017probabilistic} propose the
coupling optional P\'{o}lya tree prior, which jointly generates two
dependent random distributions through a random-partition-and-assignment
procedure similar to P\'{o}lya trees. The former paper consider both
testing hypotheses from a global point of view, while the latter takes a
local perspective on the two-sample hypothesis, detecting high resolution
local differences. \cite{bhattacharya2012nonparametric} propose a
Dirichlet process (DP) mixture model for testing whether there is a
difference in distributions between groups of observations on a manifold.
Both \cite{chen&hanson:14} and \cite{holmes2015two} consider the
two-sample testing problem, using a P\'{o}lya tree prior for the common
distribution in the null, while the model for the alternative hypothesis
assumes that the two population distributions are independent draws from
the same P\'{o}lya tree prior. Their approaches differ in the way they
specify the P\'{o}lya tree prior.
\cite{gutierrez2019} consider a related problem,  where a Bayesian
nonparametric strategy to test for differences between a control group and
several treatment regimes is proposed. \cite{pereiraetal:20} extend this
idea to testing equality of distributions of paired samples, with a model
for the joint distribution of both samples defined as a mixture of DPs
with a spike-and-slab prior specification for its base measure.

Another traditional (and fruitful) approach for modeling data arising from
a collection of groups or related studies involves the construction of
hierarchical random prior probability measures. One of the first such
examples in the BNP literature, is the well-known hierarchical DP mixtures
introduced in \cite{Teh_etal_HDP_2006}. Generalizations beyond the DP case
are currently an active area of research, as testified by a series of
recent papers dealing with various such hierarchical constructions; these
include \cite{Cam_etal_2019_HP}, \cite{argiento2019hcrm} and
\cite{bassetti2018hierarchical}. See the discussion below.

Our first contribution is the introduction of a novel class of
nonparametric priors that, just as discussed
in~\cite{Cam_etal_2018latent}, avoids the degeneracy issue of the nested
Dirichlet process (NDP) of \cite{rodriguez2008nested} that arises from the
presence of shared atoms across populations. Indeed,
\cite{Cam_etal_2018latent} showed that under the NDP, if two populations
share at least one common latent variable in the mixture model, then the
model identifies the corresponding distributions as completely equal. To
overcome the degeneracy issue, they resort to a latent nested construction
in terms of normalized random measures that adds a shared random measure
to draws from the NDP. Instead, we use a variation of the hierarchical DP
(HDP), that we term the semi-HDP, but where the baseline distribution is
itself a mixture of a DP and a non-atomic measure. We will show that this
procedure solves the degeneracy problem as well.
While relying on a different model, \cite{lijoi2020flexible} also propose to build on the HDP, combining it with the NDP, to overcome the degeneracy issue of nested processes.

Our second contribution is that the proposed model overcomes some of the
practical and applied limitations of the latent nested approach by
\cite{Cam_etal_2018latent}. As pointed out in
\cite{beraha_guglielmi_discussion}, the latent nested approach becomes
computationally burdensome in the case of $I>2$ populations. In contrast,
implementing posterior inference for  the semi-HDP prior   does not
require restrictions on $I$. We discuss in detail how to carry out
posterior inference in the context of hierarchical models based on the
semi-HDP.

A third contribution of this article is that we combine the proposed
semi-HDP prior with a random partition model that allows different
populations to be grouped in clusters that are internally homogeneous,
i.e. arising from the same distribution. See an early discussion of this
idea in the context of contingency tables in~\cite{quintana:98}.  The far
more general extension we aim for here is also useful from the applied
viewpoint of finding out which, if any, of the $I$ populations are
internally homogeneous when homogeneity of the whole set does not hold.
For the purpose of assessing global exchangeability, one may resort to
discrepancy measures~\citep{gelman1996posterior}; see also
\cite{catalano2021annals}.
In our approach, homogeneity corresponds to a point-null hypothesis about
a discrete vector parameter, as we adopt a \virgolette{larger} model for
the alternative hypothesis within which homogeneity is nested.   We
discuss the specific case of adopting Bayes factors for the proposed test
within the partial exchangeability framework. We show that the Bayes
factor for this test is immediately available, and derive some of its
theoretical properties.

The rest of this article is organized as follows.
Section~\ref{sec:assesing} gives some additional background that is
relevant for later developments, presents the semi-HDP prior
(Section~\ref{sec:model}) and, in particular, it describes a food court of
Chinese restaurants with private and shared areas metaphor
(Section~\ref{sec:restaurant}). Section~\ref{sec:theory} studies several
theoretical properties of the semi-HDP such as support, moments, the
corresponding partially exchangeable partition probability function (in a
particular case) and specially how the degeneracy issue is overcome under
this setting. Section~\ref{sec:BF} specializes the discussion to the
related issue of testing homogeneity when
$I=2$ populations are present, and we study properties of the Bayes
Factor for this test.
Section~\ref{sec:MCMC} describes a computational strategy to implement
posterior inference for the class of hierarchical models based on our
proposed semi-HDP prior. Extensive simulations, with $I=2$, $4$ and $100$
populations are presented in Section~\ref{sec:experiments}. An application
to an educational data set is discussed in Section~\ref{sec:data}. The
article concludes with a discussion in Section~\ref{sec:disc}. An appendix collects the proofs for
the theoretical results and a discussion on consistency for the Bayes Factor in the case of $I=2$
homogeneous populations. Code for posterior inference has been implemented
in \texttt{C++} and is available as part of the BayesMix
library\footnote{https://github.com/bayesmix-dev/bayesmix}.


\section{Assessing Exchangeability within a Partially Exchangeable Framework}\label{sec:assesing}

While exchangeability can be explored in more generality, for clarity of
exposition we set up our discussion in the context of continuous
univariate responses,  but extensions to, e.g. multivariate responses, can
be straightforwardly accommodated in our framework.

\subsection{A common home for exchangeability and partial
exchangeability}
\label{sec:exch}

A flexible nonparametric model for each group can be constructed by
assuming a mixture, where the mixing group-specific distribution $G_i$ is
a random discrete probability measure (r.p.m.), i.e.
\begin{equation}
\label{eq:mixtHi}
y_{ij}\mid G_i\iid  p_i(\cdot)=\int_{\Theta}k(\cdot\mid\theta)\, G_i(d\theta),\qquad
j=1,\ldots, N_i,
\end{equation}
where $k(\cdot\mid\theta)$ is a density in $\Y$ for any $\theta\in\Theta$,
and $G_i$ is, for example, a DP on $\Theta$.   Note that, with a little
abuse of notation, $p_i$ in \eqref{eq:mixtHi} and in the rest of the paper
denotes the conditional population density of group $i$ (before $p_i$
represented  the  population distribution of group $i$ in de Finetti's
theorem). In what follows, we will always assume that the parametric space
is contained in $\mathbb{R}^p$ for some positive integer $p$, and we will
always assume the Borel $\sigma$--field $\mathcal{B}(\Theta)$ of $\Theta$.
Using the well-known alternative representation of the mixture in terms of
latent variables, the previous expression is equivalent to assuming that
for any $i$,
\begin{equation}
y_{ij}\mid\theta _{ij} \ind k(\cdot\mid\theta _{ij}), \quad \theta_{ij}\mid G_i\iid G_i, \quad j=1,\ldots,N_i.
\label{eq:latent_rep}
\end{equation}
In this case, partial exchangeability of observations $(y_{ij})_{ij}$ is
equivalent to  partial exchangeability of the latent variables
$(\theta_{ij})_{ij}$. Hence exchangeability of observations
$(y_{ij})_{ij}$  is equivalent to the statement $G_1=G_2= \cdots = G_I$
with probability one.

In the next subsection we develop one of the main contributions of this
paper, namely, the construction of a prior distribution
$\pi(G_1,\ldots,G_I)$  such that there is positive prior probability that
$G_1=G_2=\cdots = G_I$, but avoiding the degeneracy issues discussed in
\cite{Cam_etal_2018latent} and that would arise if we assumed that
$(G_1,\ldots, G_I)$ were distributed as the NDP by
\cite{rodriguez2008nested}. Briefly, $(G_1,\ldots, G_I)$ is distributed as
the NDP if
\[
	G_i \mid G \iid G = \sum_{\ell=1}^{\infty} \pi_\ell \delta_{G^*_\ell},\quad i=1,\ldots,I \quad \text{and} \quad
G^*_\ell \iid Q_0= \mathcal{D}_{\gamma G_{00}},
\]
i.e., the independent atoms in $G$ are all drawn from a DP on $\Theta$,
specifically $G^*_\ell = \sum_{h=1}^{\infty} w_{h\ell}
\delta_{\theta_{h\ell}}$, with  $\theta_{h\ell}\iid G_{00}$, a probability
measure on $\Theta$, and $\alpha,\gamma>0$. The weights $(\pi_j)_j$ and $
(w_{h\ell})_h $, $\ell=1,2,\ldots$, are independently obtained from the
usual stick-breaking construction, with parameters $\alpha$ and $\gamma$,
respectively. Here $\mathcal{D}_{\gamma G_{00}}$ denotes the Dirichlet
measure, i.e. the distribution of a r.p.m. that is a DP with measure
parameter $\gamma G_{00}$. However, nesting discrete random probability
measures produces degeneracy to the exchangeable case. As mentioned in
Section~\ref{sect:intro}, \cite{Cam_etal_2018latent} showed that the
posterior distribution
degenerates to the exchangeable case whenever a shared component is
detected, i.e., the NDP does not allow for sharing clusters among
non-homogeneous populations. The problem is shown to affect any
construction that uses nesting, and not just the NDP.

To overcome the degeneracy issue, while retaining flexibility,
\cite{Cam_etal_2018latent} proposed the so-called Latent Nested
Nonparametric priors.  These models involve a shared random measure that
is added to the draws from a Nested Random Measure, hence accommodating
for shared atoms. See also the discussion by
\cite{beraha_guglielmi_discussion}. There are two key ideas in their
model: ($i$) nesting discrete random probability measures as in the case
of the  NDP, and  ($ii$) contaminating the population distributions with a
common component as in \cite{muller_etal2004} and also,
\cite{Lijoi_etal_Ber14}. The contamination aspect of the model yields
dependence among population-specific random probability measures, and
avoids the degeneracy issue pointed out by the authors, while the former
accounts for testing homogeneity in multiple-sample problems. Their
approach, however, becomes computationally burdensome in the case of $I>2$
populations, and it is not clear how to extend their construction to allow
for the desired additional analysis, i.e. assessing which, if any, of the
$I$ populations are internally homogeneous when homogeneity of the whole
set does not hold.


\subsection{The Model}

\label{sec:model}

We present now a hierarchical model that allows us to assess homogeneity,
while avoiding the undesired degeneracy issues and which further enables
us to construct a grouping of populations that are internally homogeneous.
To do so we create a hierarchical representation of distributions that
emulates the behavior arising from an exchangeable partition probability
function~\cite[EPPF;][]{pitman2006combinatorial} such as the P\'olya urn.
But the main difference with previous proposals to overcome degeneracy is
that we now allow for different populations to arise from the same
distribution,  while simultaneously incorporating an additional mechanism
for populations to explicitly differ from each other.

Denote $[I]=\{1,\ldots,I\}$. A partition $S_1,\ldots,S_k$ of $[I]$ can be
described by  cluster assignment indicators $\bm{c}=(c_1,\ldots,c_I)$ with
$c_i=\ell$ iff $i\in S_\ell$. Assume this partition arises from a given
EPPF. We introduce the following model for the latent variables in a
mixture model such as \eqref{eq:latent_rep}. Let $\bm y_i:=(y_{i1},\ldots,
y_{i N_i})$, for $i=1,\ldots,I$. We assume that $\bm y_1,\ldots,\bm y_I$,
given all the population distributions $F_1,\ldots,F_I$ are independent,
and furthermore arising from
\begin{align}
	y_{ij} \mid F_1, \dots, F_I, \bm c & \iid \int_{\Theta}k(\cdot\mid\theta)\, F_{c_i}(d\theta), \   j=1,\ldots, N_i, \ \textrm{ for all } i
	\label{eq:hdp_lik} \\
	\bm c & \sim \pi_c(c_1, \dots, c_I)
	\label{eq:prior_c}\\
	F_1, \dots F_I \mid \Ptilde &\iid \mathcal{D}_{\alpha \Ptilde} \label{eq:iid_DP}  \\
	\Ptilde &= \kappa G_0 + (1-\kappa) \Gtilde
	\label{eq:hdp_p0} \\
	\Gtilde &\sim \mathcal{D}_{\gamma G_{00}}
	\label{eq:hdp_g0}\\
	\kappa & \sim Beta(a_\kappa, b_\kappa),
	\label{eq:wprior}
\end{align}
where $\alpha,\gamma>0$.  Thus the role of the population mixing
distribution $G_i$ in~\eqref{eq:mixtHi} -- or, equivalently, in
\eqref{eq:latent_rep} -- is now played by $F_{c_i}$. Observe that $F_1,
\ldots, F_I$ in~\eqref{eq:iid_DP} play a role similar to the cluster
specific parameters in more standard mixture models. Consider for example
a case where $I=4$ and $\bm c = (1, 2, 3, 1)$. Under the above setting,
$F_1, F_2, F_3$ define a model for three different distributions, so that
populations 1 and 4 share a common mixing distribution, and $F_4$ is never
employed.

Equation \eqref{eq:iid_DP} means that conditionally on $\Gtilde$ each
$F_k$ is an  independent draw from a DP prior with mean parameter
$\Ptilde$ (and total mass $\alpha$), i.e. $F_k$ is a discrete r.p.m. on
$\Theta \subset \R^p$ for some positive integer $p$, with $F_k = \sum_{h
\geq 1} w_{kh} \delta_{\theta^*_{kh}}$ where for any $k$ the weights are
independently generated from a stick-breaking process, $\{w_{kh}\}_h\iid
SB(\alpha)$, i.e.
\[
w_{k1} = \beta_{k1},\quad \ w_{kh}=\beta_{ih} \prod_{j=1}^{h-1}(1 - \beta_{kj}) \ \
\mbox{for $h=2,3,\ldots$}, \quad \beta_{ij}\iid Beta(1,\alpha),
\]
and $\{\theta^*_{kh}\}_h$, $\{\beta_{kh}\}_h$ are independent, with
$\theta^*_{kh} \mid \Ptilde \iid \Ptilde$. We assume the centering measure
$\Ptilde$ in \eqref{eq:hdp_p0} to be a \textit{contaminated draw}
$\Gtilde$ from a  DP prior, with centering measure $G_{00}$, with a fixed
probability measure $G_0$. Both $G_0$ and $G_{00}$ are assumed to be
absolutely continuous (and hence non-atomic) probability measures defined
on $\left( \Theta, \mathcal B(\Theta)\right)$.

By \eqref{eq:hdp_g0}, $\Gtilde = \sum_{h \geq 1} p_h \delta_{\tau_h}$,
where $\{p_h\}_h\sim SB(\gamma)$, $\tau_h\iid G_{00}$ are independent
weights and location points. The model definition is completed by
specifying  $\pi_c(c_1, \dots, c_I)$. We assume that the $c_i$'s are
(conditionally) i.i.d. draws from a categorical distribution on $[I]$ with
weights $\bm \omega = (\omega_1, \dots, \omega_I)$, i.e. $c_i \mid \bm
\omega \iid Cat([I];\, \bm \omega)$, where the elements of $\bm\omega$ are
non-negative and constrained to add up to 1. A convenient prior for $\bm
\omega$ is a finite dimensional Dirichlet distribution with parameter $\bm
\eta = (\eta_1, \dots \eta_I)$. Observe that distributions $F_{c_1},
\ldots, F_{c_I}$ allow us to cluster populations, so that there are at
most $I$ clusters and consequently $F_1, \ldots, F_I$ are all of the
cluster distributions that ever need to be considered.

We say that a vector of random probability measures $(F_1, \dots, F_I)$
has the semi-hierar\-chi\-cal Dirichlet process (semi-HDP) distribution if
\eqref{eq:iid_DP}-\eqref{eq:hdp_g0} hold, and we write $(F_1, \dots, F_I)
\sim semiHDP(\allowbreak \alpha, \gamma, \kappa, G_0, G_{00})$. It is straightforward
to prove that, conditional on $\kappa$ and eventual hyperparameters in
$G_0$ and $G_{00}$,
the expectation of any $F_i$ is $\kappa G_0 + (1 - \kappa) G_{00}$
which further reduces to $G_{00}$ if $G_0 = G_{00}$. Note that $(F_1,
\dots, F_I) \sim semiHDP(\alpha, \gamma, \kappa, G_0, G_{00})$ defines an
exchangeable prior over a vector of random probability measures.

We note several immediate yet interesting properties of the model. First,
note that if $\kappa=1$ in \eqref{eq:hdp_p0}, then all the atoms and
weights in the representation of the $F_i$'s are independent and different
with probability one, since the beta distribution and $G_0$ are absolutely
continuous. If $\kappa = 0$, then our prior
\eqref{eq:iid_DP}-\eqref{eq:hdp_g0} coincides with  the Hierarchical
Dirichlet Process in \cite{Teh_etal_HDP_2006}. Since $\Gtilde = \sum_{h
\geq 1} p_h \delta_{\tau_h}$, then, with positive probability, we have
$\theta^*_{kh} = \theta^*_{k'm} = \tau_\ell$ for $k \neq k'$, i.e. all the
$F_k$'s share the same atoms in the stick-breaking representation of
$\Gtilde$. However, even when $\kappa=0$, $F_k \neq F_j$ with probability
one, as the weights $\{w_{kh}\}_h$ and $\{w_{jh}\}_h$ are different, since
they are built from (conditionally) independent stick-breaking priors.
This is precisely the feature that allows us to circumvent the degeneracy
problem.

Second, our model introduces a vector parameter $\bm c$, which assists
selecting each population distribution from the finite set
$F_1,\ldots,F_I$, in turn assumed to arise from the semi-HDP prior
\eqref{eq:iid_DP}-\eqref{eq:hdp_g0}. The former allows two different
populations to have the same distribution (or mixing measure) with
positive probability, while the latter allows to overcome the degeneracy
issue while retaining exchangeability. Indeed, as noted above, $F_i$ and
$F_j$ may share atoms. The atoms in common arise from the atomicity of the
base measure
and we let the atomic component of the base measure to be a draw from a
DP. The result is a very flexible model, that on one hand is particularly
well-suited for problems such as density estimation, and on the other, can
be used to construct clusters of the $I$ populations, as desired.

\subsection{A restaurant representation} \label{sec:restaurant}

To better understand the cluster allocation under model
\eqref{eq:hdp_lik}-\eqref{eq:hdp_g0}, we rewrite  \eqref{eq:hdp_lik}
introducing the latent variables $\{\theta_{ij}\}$ as follows
\begin{align}
    y_{ij} \mid F_1, \ldots F_I, \bm c , \theta_{ij} &\ind k(\cdot \mid \theta_{ij})  
    \label{eq:hdp_lik_lat}\\
    \theta_{i1}, \ldots \theta_{iN_i} \mid   F_1, \ldots F_I, \bm c & \iid F_{c_i} 
    \label{eq:hdp_lik_lat2}
\end{align}
and $\{\theta_{i\ell}\}_\ell \perp \{\theta_{jm}\}_m$ for $i\ne j$,
conditionally on $F_1, \ldots, F_I$.

We first derive   the conditional law of the $\theta_{ij}$'s under
\eqref{eq:hdp_lik_lat} - \eqref{eq:hdp_lik_lat2}, and
\eqref{eq:prior_c}-\eqref{eq:hdp_p0}, given $\Gtilde$. All customers of
group $i$ enter restaurant $r$ (such that $c_i = r$). If group $i$ is the
first group entering restaurant $r$, then the usual Chinese Restaurant
metaphor applies. Instead, let us imagine that group $i$ is the last group
entering restaurant $r$ among those such that $c_m=r$. Upon entering the
restaurant, the customer is presented with the usual Chinese Restaurant
Process (CRP), so that
\begin{equation}
\theta_{ij} \mid \bm c, \{\theta_{m k}, \ \forall m: c_m=c_i=r \}, \theta_{i1}, \ldots, \theta_{ij-1},
\Gtilde \\
\sim \sum_{\ell=1}^{H_r} \frac{n_{r\ell}}{\alpha + n_{r \bdot}} \delta_{\theta^*_{r\ell}} + \frac{\alpha}{\alpha + n_{r \bdot}} \Ptilde,
\label{eq:crp_restaurant}
\end{equation}
that is the CRP when considering all the groups entering restaurant $r$ as
a single group. Here $H_r$ denotes the number of tables in restaurant $r$,
and $n_{r\ell}$ is the number of customers who entered from restaurant $r$
and are seating at table $\ell$. Moreover, note that $\theta^*_{r\ell}
\mid \Gtilde \iid \Ptilde$, so that, as in the HDP, there might be ties
among the $\theta^*_{r\ell}$ also when keeping $r$ fixed.  This is an
important observation as the fact that there might be ties for different
values of $r \neq r'$ instead, is exactly what let us avoid the degeneracy
to the exchangeable case.  Note that \eqref{eq:crp_restaurant} holds also
for $\theta_{i1}$, i.e. the first customer in group $i$. In the following,
we will use \emph{clusters} or \emph{tables} interchangeably. However,
note that, unlike traditional CRPs, the number of \emph{clusters} does not
coincide with the number of unique values in a sample. This point is
clarified in \cite{argiento2019hcrm}, who introduce the notion of
$\ell$--cluster, which is essentially the \emph{table} in our restaurant
metaphor.

\begin{figure}[t]
\centering
\includegraphics[width=\linewidth]{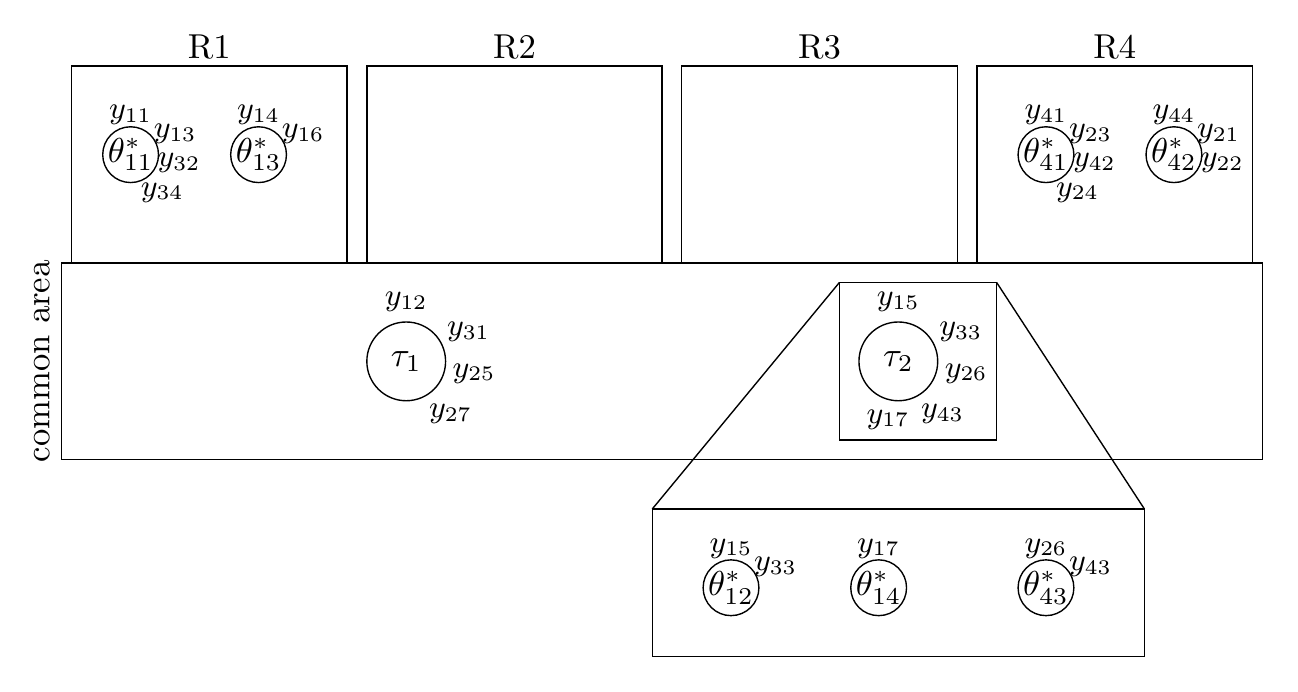}
\caption{Restaurant representation of the semi-HDP allocation. In the image, $\bm c
= (1, 4, 1, 4)$ so that groups one and three enter in restaurant R1 while
groups two and four enter in restaurant R4. In the \virgolette{common
area} two tables are represented, $\tau_1$ and $\tau_2$.
\virgolette{Zooming} into $\tau_2$ shows that there are three different
$\theta^*$'s associated to the value $\tau_2$, namely $\theta^*_{12},
\theta^*_{13}$ and $\theta^*_{43}$. The first two originate from R1,
showing that it is possible to have ties among the $\theta^*$'s even
inside the same restaurant, while the table labeled $\theta^*_{43}$ shows
that it is possible to have ties across different restaurants.}
\label{fig:restaurant}
\end{figure}

Observe from \eqref{eq:crp_restaurant} that when a new cluster is created,
its label is sampled from $\Ptilde$.
In practice, we augment the parameter space with a new binary latent
variable for each cluster, namely $h_{r\ell}$, with
$h_{r\ell}\iid\text{Bernoulli}(\kappa)$, so that
\[
	\theta^*_{r\ell} \mid h_{r\ell}=1 \ \sim G_0 \qquad \text{ and } \qquad
	\theta^*_{r\ell} \mid h_{r\ell}=0, \Gtilde \ \sim \Gtilde.
\]
Upon conditioning on $\{h_{r\ell}\}$ it is straightforward to integrate
out $\Gtilde$. Indeed, we can write the joint distribution of
$\{\theta^*_{r\ell}, \ \forall r \ \forall \ell\}$, conditional on
$\{h_{r\ell}\}$ as
$$    \{\theta^*_{r\ell}\} \mid \{h_{r\ell}\}, \Gtilde \sim \prod_{r, \ell}G_0(d\theta^*_{r\ell})^{h_{r\ell}}
    \ \prod_{r, \ell}\Gtilde(d\theta^*_{r\ell})^{1 - h_{r\ell}}.$$
Hence we see that $\{\theta^*_{r\ell}, \ \forall r \ \forall \ell:
h_{r\ell} = 0\}$ is a conditionally i.i.d sample from $\Gtilde$ (given all
the $h_{rl}$'s and $\Gtilde$), so that we can write:
\begin{equation}
	\theta^*_{r\ell} \mid h_{r\ell} = 0, \{\theta^*_{ij}: h_{ij} = 0\} \sim \sum_{k=1}^{H_0}
\frac{m_{\bdot k}}{m_{\bdot \bdot} + \gamma} \delta_{\tau_k} + \frac{\gamma}{m_{\bdot \bdot} + \gamma} G_{00}
	\label{eq:crp_shared}
\end{equation}
and $\tau_k \iid G_{00}$, where $H_0$ denotes the number of tables in the
common area in Figure~\ref{fig:restaurant}, and $m_{rk}$ denotes the
cardinality of the set $\{\theta^*_{r\ell} : \ \theta^*_{r\ell} = \tau_k
\}$. The dot subindex denotes summation over the corresponding subindex
values. Hence, conditioning on all the $(r, \ell)$ such that $h_{r\ell} =
0$, with $r$ corresponding to a non-empty restaurant, we recover the
Chinese Restaurant Franchise (CRF) that describes the HDP.

We can describe the previously discussed clustering structure in terms of
a restaurant metaphor as the \virgolette{food court of Chinese restaurants
with private and shared areas}. Here, the $\theta^*_{r\ell}$ correspond to
the tables and $\theta_{ij}$ to the customers. Moreover, a dish is
associated to each table. Dishes are represented by the various
$\theta^*_{r\ell}$'s . There is one big common area where tables are
shared among all the restaurants and $I$ additional ``private''  small
rooms, one per restaurant, as seen in Figure~\ref{fig:restaurant}. The
common area accommodates tables arising from the HDP, i.e. those tables
such that $\tau_k \iid G_{00}$, while the small rooms host those tables
associated to non empty restaurants, such that $\theta^*_{r\ell} \mid
h_{r\ell} = 1 \iid G_0$. All the customers of group $i$ enter restaurant
$r$ (such that $c_i = r$). Upon entering the restaurant, a customer is
presented with a menu. The $H_r$ dishes in the menu are the
$\theta^*_{r\ell}$'s, and because $\theta^*_{r\ell} \iid \Ptilde$, there
might be repeated dishes; see~\eqref{eq:crp_restaurant}. The customer
either chooses one of the dishes in the menu, with probability
proportional to the number of customers who entered the same restaurant
and chose that dish, or a new dish (that is not included in the menu yet)
with probability proportional to $\alpha$; again,
see~\eqref{eq:crp_restaurant}. If the latter option is chosen, with
probability $\kappa$ a new table is created in the restaurant-specific
area, $H_r$ is incremented by one and a new dish $\theta^*_{rH_r+1}$ is
drawn from $G_0$. With probability $1-\kappa$ instead, the customer is
directed to the shared area, where (s)he chooses to seat in one of the
occupied tables with a probability proportional to $m_{\cdot k}$, i.e. the
number of items in the menus (from all the restaurants) that are equal to
dish $\tau_k$,
or seats at a new table with a probability proportional to $\gamma$, as
seen from~\eqref{eq:crp_shared}. We point out that the choice of table in
this case is made without any knowledge of which restaurant the dishes
came from. Moreover, if the customer chooses to sit at a new table, we
increment $H_0$ by one and draw $\tau_{H_0 + 1} \sim G_{00}$; we also
increment $H_r$ by one and set $\theta^*_{rH_r+1} = \tau_{H_0 + 1}$.
Observe that in the original CRF metaphor, it is not the tables that are
shared across restaurants, but rather the dishes. In our metaphor instead,
we group together all the tables corresponding to the same $\tau_h$ and
place them in the shared area. This is somewhat reminiscent of the direct
sampler scheme for the HDP. Nevertheless, observe that the bookkeeping of
the $m_{rk}$'s is still needed. To exemplify this, in
Figure~\ref{fig:restaurant} we report a \virgolette{zoom} on a particular
shared table $\tau$, showing that the $\theta^*$'s associated to that
table are still present in our metaphor, but can be collapsed into a
single shared table when it is convenient.

\section{Theoretical properties of the semi-HDP prior}\label{sec:theory}
Here we develop additional properties of the proposed prior model. In
particular, we study the topological support of the semi-HDP and show how
exactly the degeneracy issue is resolved by studying the induced joint
random partition model on the $I$ populations.

\subsection{Support and moments}
An essential requirement of nonparametric priors is that they should have
large topological support; see \cite{Fergus73}. Let us denote by $\pi_{\bm
G}$ the probability measure on $\mathbb{P}_{\Theta}^I$ corresponding to
the prior distribution $\pi(G_1,\ldots,G_I)$ of
the random vector $(G_1,\ldots,G_I)$ specified in
\eqref{eq:prior_c}--\eqref{eq:hdp_g0}, with $G_i=F_{c_i}$; see
\eqref{eq:mixtHi}. We show here that the prior probability measure
$\pi_{\bm G}$ has full weak support, i.e. given any point $\bm
g=(g_1,\ldots, g_I)$ in $\mathbb{P}_{\Theta}^I$, $\pi_{\bm G}$ gives
positive mass to any weak neighborhood $\mathcal{U}(\bm g;\epsilon)$ of
$\bm g$, of diameter $\epsilon$.

\begin{prop}[Full Weak Support]\label{prop:weak_supp} Let
$\pi_{\bm G}(g_1\ldots,g_I)$ be the prior probability measure on
$\mathbb{P}_{\Theta}^I$ defined by
\eqref{eq:prior_c}--\eqref{eq:hdp_g0}.
\begin{description}
\item[(a)] If $G_0$ in \eqref{eq:hdp_p0} has full support on $\Theta$
    and $0 < \kappa \leq 1$, then $\pi_{\bm G}(g_1\ldots,g_I)$  
    has full weak support.
\item[(b)] If $\kappa=0$ and $G_{00}$ in \eqref{eq:hdp_g0} has full
    support, then $\pi_{\bm G}(g_1\ldots,g_I)$  
    has full weak support.
\end{description}

\noindent Proof: see the Appendix, Section~\ref{sec:s_proofs}.
\end{prop}

It is straightforward to show that in case where $\pi_c(c_1,\ldots,c_I)$
is exchangeable and $P(c_i = \ell) = \omega_\ell$ for $\ell = 1, \ldots,
I$ then  \eqref{eq:hdp_lik}--\eqref{eq:hdp_g0} becomes, after
marginalizing with respect to $\bm c$,
\begin{align*}
y_{ij}\mid F_1, \dots ,F_I  &\iid \sum_{\bm c} \int_{\Theta}k(\cdot\mid\theta)\, F_{c_i}(d\theta) \pi_c(c_1,\ldots,c_I) = \sum_{\ell=1}^I \omega_\ell \int_{\Theta}k(\cdot\mid\theta)\, dF_{\ell}(\theta).
\end{align*}
In this case, the conditional marginal distribution of each observation
can be expressed as a finite mixture of mixtures of the density
$k(\cdot\mid\theta)$ with respect to each of the random measures
$F_1,\dots,F_I$, i.e. a finite mixture of Bayesian nonparametric mixtures.

We have mentioned above that in the case in which $G_{00} = G_0$ in
Equations \eqref{eq:hdp_p0} - \eqref{eq:hdp_g0}, the marginal law of $F_i$
is $G_{0}$, and equivalently, for each $A \in \mathcal{B}(\Theta)$,
$\mathbb{E}[F_i(A)] = G_{0}(A)$  for any $i$. In this case, the covariance
between $F_1$ and $F_2$ is given by
$$\cov \left(F_1(A), F_2(B) \right) =  \frac{(1 - \kappa)^2}{1 + \gamma}
 \left( G_0(A \cap B) - G_0(A) G_0(B) \right).
$$
See the Appendix, Section~\ref{sec:s_proofs}, for the
proof of these formulas. Note that, in the case of Hierarchical Normalized
Completely Random Measures, and hence in the HDP, the covariance between
$F_1$ and $F_2$ depends exclusively on the intensity of the random measure
governing $\Gtilde$ (in the case of the DP the dependence is on $\gamma$).
For instance, see \cite{argiento2019hcrm}, Equation (5) in the
Supplementary Material. Instead, in the Semi-HDP, an additional parameter
can be used to tune such covariance: the weight $\kappa$. Indeed, as
$\kappa$ approaches $1$, the two measures become more and more
uncorrelated, the limiting case being full independence as discussed at
the end of Section~\ref{sec:model}. In the Appendix,
Section~\ref{sec:s_proofs}, we also report an expression for the higher
moments of $F_i(A)$ for any $i$.

\subsection{Degeneracy and marginal law}
We now formalize the intuition given in Section~\ref{sec:restaurant} and
show that our model, as defined in \eqref{eq:hdp_lik}-\eqref{eq:hdp_g0},
does not incur in the degeneracy issue described by
\cite{Cam_etal_2018latent}. The degeneracy of a nested nonparametric
model refers to the following situation: if there are shared values (or
atoms in the corresponding mixture model) across any two populations, then
the posterior of these population/random probabilities degenerates,
forcing homogeneity across the corresponding samples. See also the
discussion in \cite{beraha_guglielmi_discussion}.

From the food court metaphor described above, it is straightforward to see
that degeneracy is avoided if two customers sit in the same table (of the
common area) with positive probability, conditioning on the event that
they entered from two different restaurants.

To see that this is so for the proposed model, let us consider the case $I
= 2$ and $\theta_{i1} \mid F_1, F_2, \bm c=(1,2) \sim F_i$, for $i=1,2$.
Marginalizing out $(F_1, F_2)$, this is equivalent to $\theta_{11},
\theta_{21} \mid \Gtilde, \ \bm c=(1,2) \ \iid w G_0 + (1-w) \Gtilde$. Now, since $G_0$ is
absolutely continuous, $\{\theta_{11} = \theta_{21}\}$ if and only if
\begin{enumerate*}[label=(\roman*)]
\item $\theta_{11}$ and $\theta_{21}$ are sampled i.i.d. from $\Gtilde$;
    and
\item we have a tie (which arises from the P\'olya-urn scheme), i.e.
    $\theta_{21} = \tau_1 = \theta_{11}$ and $\tau_1 \sim G_{00}$.
\end{enumerate*}
This means that   $\theta_{11}$, the first customer, sits in a table of
the common area, an event that happens with probability $1-\kappa$ since
she is the first one in the whole system, and $\theta_{21}$ decides to sit
in the common area (with probability $1-\kappa$) and subsequently decides
to sit at the same table of $\theta_{11}$ (which happens with probability
$\frac{1}{\gamma +1})$. Summing up we have that $	p(\theta_{11} =
\theta_{21} \mid \bm c=(1,2))  = (1 - \kappa)^2 / (1 + \gamma) $ which is
strictly positive if $\kappa < 1$. Hence, by Bayes' rule, we have that
\[
    P(c_1 \neq c_2 \mid \theta_{11} = \theta_{21}) = \frac{
    P(\theta_{11} = \theta_{21} \mid c_1 \neq c_2)
    P(c_1 \neq c_2)}{\sum_{i,j} P(\theta_{11} = \theta_{21} \mid \bm c=(i,j)) P(\bm c = (i, j) )} > 0.
\]
%
Moreover, when $\kappa=1$ we find the same degeneracy issue described in
\cite{Cam_etal_2018latent}, as proved in Proposition \ref{prop:degeneracy}
below.

To get a more in-depth look at these issues, we follow
\cite{Cam_etal_2018latent} and study properties of the partially
exchangeable partition probability function (pEPPF) induced by our model,
which we define in the special case of $I=2$. Consider a sample $\bm
\theta=(\bm \theta_1,\bm \theta_2)$ of size $N = N_1 + N_2$ from model
\eqref{eq:hdp_lik_lat2}, together with
\eqref{eq:prior_c}-\eqref{eq:hdp_g0} for $I=2$ populations; let $k = k_1 +
k_2 + k_0$ the number of unique values in the samples, with $k_1$ ($k_2$)
unique values specific to group 1 (2) and $k_0$ shared between the groups.
Call $\bm n_i$ the frequencies of the $k_i$ unique values in group $i$ and
$\bm q_i$ the frequencies of the $k_0$ shared values in group $i$; this is
the same notation as in \cite{Cam_etal_2018latent}, Section~2.2. The pEPPF
is defined as
\[
\Pi^{N}_k(\bm n_1, \bm n_2, \bm q_1, \bm q_2 \mid \bm c=(\ell, m)) =
\int_{{\Theta}^k} \mathbb{E} \left[ \prod_{j=1}^{k_1}
F_\ell^{n_{1j}}(d\theta^*_{1j}) \prod_{j=1}^{k_2} F_m^{n_{2j}}(d\theta^*_{2j})
\prod_{j=1}^{k_0} F_\ell^{q_{1j}}(d\tau_{j}) F_m^{q_{2j}}(d\tau_{j})  \right]
\]

\begin{prop}\label{prop:degeneracy}
Let $\kappa$ in \eqref{eq:hdp_p0} be equal to 1, let $\pi_1 = P(c_1=c_2)$,
then the pEPPF $\Pi^{(N)}_k  (\bm n_1, \bm n_2, \bm q_1, \bm q_2)$ can be
expressed as:
\begin{multline}\label{eq:peppf_deg}
	\Pi^{(N)}_k  (\bm n_1, \bm n_2, \bm q_1, \bm q_2) = \pi_1
\Phi_k^{(N)}(\bm n_1, \bm n_2, \bm q_1 + \bm q_2) \\ + (1-\pi_1) \Phi_{k_0
+ k_1}^{(N_1)} (\bm n_1, \bm q_1) \Phi_{k_0 + k_1}^{(N_2)} (\bm n_2, \bm
q_2) I(k_0 = 0)
\end{multline}
where \[
 \Phi_k^{(N)}(\bm n_1, \bm n_2, \bm q_1 + \bm q_2) = \frac{\alpha^{k_1 + k_2 + k_0} \Gamma(\alpha)}{\Gamma(\alpha + N)}
 \prod_{j=1}^{k_1} \Gamma(n_{1j}) \prod_{j=1}^{k_2} \Gamma(n_{2j}) \prod_{j=1}^{k_0} \Gamma(q_{1j} + q_{2j})
\]
is the EPPF of the fully exchangeable case, and
\[
 \Phi_{k_0 + k_i}^{(N_i)} (\bm n_i, \bm q_i) = \frac{\alpha^{k_i + k_0} \Gamma(\alpha)}{\Gamma(\alpha + N_i)} \prod_{j=1}^{k_i} \Gamma(n_{ij}) \prod_{j=1}^{k_0} \Gamma(q_{ij}),  \ i=1,2
\]
is the marginal EPPF for the individual group $i$.

\noindent
Proof: see the Appendix, Section~\ref{sec:s_proofs}.
\end{prop}
This result shows that a suitable prior for $\kappa$ requires assigning
zero probability to the event $\kappa=1$. The assumption in
\eqref{eq:wprior} trivially satisfies this requirement.

Finally, we consider the marginal law of a sequence of vectors $(\bm
\theta_1, \ldots, \bm \theta_I)$, $\bm \theta_\ell = (\theta_{\ell 1},
\ldots \theta_{\ell N_l})$ from model
\eqref{eq:hdp_lik}-\eqref{eq:hdp_g0}. Let us first derive the marginal law
conditioning on $\bm c$, as the full marginal law will be the mixture of
these conditional laws over all the possible values of $\bm c$.
\begin{prop}\label{prop:marginal_law}
The marginal law of a sequence of vectors $(\bm \theta_1, \ldots, \bm
\theta_I)$, $\bm \theta_\ell = (\theta_{\ell 1}, \ldots \theta_{\ell
N_\ell})$ from model \eqref{eq:hdp_lik}-\eqref{eq:hdp_g0}, conditional to
$\bm c$ is
\begin{equation}
	\prod_{i=1}^{R(\bm c)} eppf(\bm n_{r_i}; \alpha) \sum_{\bm h \in \{0, 1\}^L} p(\bm h) \prod_{\ell=1}^L G_0(d\theta^*_\ell)^{h_\ell}
\times eppf(\bm m_{r_i} \mid \bm h; \gamma) \prod_{k=1}^{M} G_{00}(d \theta^{**}_k). \label{eq:marg}
\end{equation}
Here, $\{\theta^*_\ell\}_{\ell=1}^L = \{\theta^*_{11}, \ldots,
\theta^*_{IH_I}\}$ is a sequence representing all the tables in the
process, obtained by concatenating the tables in each restaurant.
Moreover,
$R(\bm c)$ is  the number of unique values in $\bm {c}$, i.e. the number
of non-empty restaurants, $\bm n_{r_i}$
is the vector of $\ell$-cluster sizes for restaurant $r_i$, $\bm m_{r_i}$
is the vector of the cluster sizes of the $\theta^*_{\ell}$ such that
$h_\ell = 0$ and $\theta^{**}_{k}$ are the unique values among such
$\theta^*_{\ell}$, where  \virgolette{$eppf$} denotes the the distribution
of the partition induced by the table assignment procedure in the food
court of Chinese restaurants described in Section~\ref{sec:restaurant}.

\noindent
Proof: see the Appendix, Section~\ref{sec:s_proofs}.
\end{prop}
The marginal law of $(\bm \theta_1, \ldots, \bm \theta_I)$ is then
\[
	\Law (d\bm \theta_1, \ldots, d\bm \theta_I) = \sum_{\bm c} \Law(d\bm \theta_1, \ldots, d\bm \theta_I \mid \bm c) \pi(\bm c)
\]
where $\Law(d\bm \theta_1, \ldots, d\bm \theta_I \mid \bm c)$ is given in \eqref{eq:marg}.

Observe that in Proposition \ref{prop:degeneracy} we denoted by $\Phi$ the
EPPF, while in \eqref{eq:marg} we use notation \virgolette{$eppf$}. This
is to remark that these objects are inherently different: $\Phi$ is the
EPPF of the partition of \emph{unique} values in the sample, while $eppf$
here is the EPPF of the tables, or $\ell$--clusters, induced by the table
assignment procedure described in Section~\ref{sec:restaurant}. Hence,
from a sample $\bm \theta$ one can recover $\bm n_1, \bm n_2, \bm q_1, \bm
q_2$ in \eqref{eq:peppf_deg} but not $\bm n_{r_i}$ in \eqref{eq:marg}.

\subsection{Some results on the Bayes factor for testing homogeneity}
\label{sec:BF}

We consider now testing for homogeneity within the proposed partial
exchangeability framework. As a byproduct of the assumed model, the
corresponding Bayes factor is immediately available. For example, if one
wanted to test whether populations $i$ and $j$ were homogeneous, it would
suffice to compute the Bayes factor for the test
\begin{equation}\label{eq:bf_ci_cj}
    H_0: c_i = c_j \quad \text{vs.} \quad H_1: c_i \neq c_j
\end{equation}
which  can be straightforwardly estimated from the output of the posterior
simulation  algorithm that will be presented later on. Note that these
\virgolette{pairwise} homogeneity tests are not the only object of
interest that we can tackle within our framework. Indeed it is possible to
test any possible combination of $\bm c$ against an alternative.

These tests admit an equivalent representation in terms of a model
selection problem; for example in the case of $I=2$ populations, we can
rewrite \eqref{eq:bf_ci_cj}, for $i=1$ and $j=2$, as a model selection
test for $M_1$ against $M_2$, where
$$M_1:\  y_{11}, \ldots, y_{1N_1}, y_{21}, \ldots, y_{2N_2} \mid F_1 \iid \int_{\Theta} k(\cdot \mid \theta) F(d\theta),
\quad F_1 \sim semiHDP(\alpha, \gamma, \kappa, G_0, G_{00})$$
and
$$M_2: \ y_{i1}, \ldots, y_{iN_i}, \mid F_i \iid \int_{\Theta} k(\cdot \mid \theta) F_i(d\theta),\,i=1,2,\quad
F_1, F_2 \sim semiHDP(\alpha, \gamma, \kappa, G_0, G_{00}). $$
In this case
\begin{equation*}
BF_{12}:=BF_{12}(y_{11}, \ldots, y_{1N_1}, y_{21}, \ldots, y_{2N_2})= \frac{m_{M_1}(y_{11}, \ldots, y_{1N_1}, y_{21}, \ldots, y_{2N_2})}{m_{M_2}(y_{11}, \ldots, y_{1N_1}, y_{21}, \ldots, y_{2N_2})},
\label{eq:BFdef}
\end{equation*}
where $m_{M_i}$ denotes the marginal law of the data under model $M_i$,
$i=1,2$, defined above. Asymptotic properties of Bayes factors have been
discussed by several authors. We refer to \cite{walker2004priors},
\cite{ghosal2008nonparametric} for a more detailed discussion and to
\cite{chib2016bayes} for a recent survey on the topic.
\cite{chatterjee2020short} is a recent and solid contribution to the
almost sure convergence of Bayes factor in the general set-up that
includes dependent data, i.e. beyond the usual i.i.d. context.

In words, our approach can be described as follows.  When the data are
assumed to be exchangeable, we assume that both samples are generated
i.i.d from a distribution $P_0$ with density $p_0$.  If the data are
instead assumed to be partially exchangeable, then we consider the first
population to be generated i.i.d from a certain $P_0$ with density $p_0$,
while the second one is generated from $Q_0$ with density $q_0$, with $P_0
\neq Q_0$
and independence holds across populations. The Bayes factor for
comparing $M_1$ against $M_2$ is thus consistent if:\\
  ($i$) $BF_{12} \rightarrow +\infty$ $P_0^\infty$--a.s. when $N_1,N_2\rightarrow +\infty$ if the groups are truly
homogeneous,  and ($ii$) $BF_{12} \rightarrow 0$ $(P_0 \otimes
Q_0)^\infty$--a.s. when $N_1,N_2\rightarrow +\infty$  if the groups are
not homogeneous.

The two scenarios must be checked separately. In the latter case,
consistency of the Bayes factor can be proved by arguing that only model
$M_2$ satisfies the so-called Kullback-Leibler property, so that
consistency is ensured by the theory in  \cite{walker2004priors}. We
summarize this result in the following proposition.
\begin{prop}\label{prop:consistency}
Assume that $y_{11}, \ldots, y_{1N_1} \iid P_0$, $y_{21}, \ldots, y_{2N_2}
\iid Q_0$, $P_0 \neq Q_0$, and that $\{y_{1i}\}$ and  $\{y_{2j}\}$ are independent.
Assume that $P_0$ and $Q_0$ are absolutely continuous measures with probability density
functions $p_0$ and $q_0$ respectively. Then, under conditions $B1$-$B9$ in
\cite{wu2008kullback}, $BF_{12} \rightarrow 0$ as $N_1,N_2\rightarrow
+\infty$.

\noindent
Proof: see the Appendix, Section~\ref{sec:s_proofs}.
\end{prop}
Observe that, out of the nine conditions $B1$-$B9$, we have that $B1-B3$,
$B7$ and $B9$ involve regularity conditions of the kernel
$k(\cdot|\theta)$. These are satisfied if the kernel is, for example,
univariate Gaussian with parameters $\theta=(\mu, \sigma^2)$. Conditions
$B4 - B6$ involve regularity of the true data generating density, which
are usually satisfied in practice. Condition $B8$ requires that the mixing
measure has full weak support, already proved in
Proposition~\ref{prop:weak_supp}.

On the other hand, when $p_0 = q_0$, consistency of the Bayes factor would
require $BF_{12} \rightarrow +\infty$. This is a result we have not been
able to prove so far. The Appendix,
Section~\ref{sec:s_bf} discusses the relevant issues arising when trying
to prove the consistency in this setting; we just report here that the key
missing condition is an upper bound of the prior mass of $M_2$. The lack
of such bounds for general nonparametric models is well known in the
literature, and not specific to our case, as it is shared, for instance,
by \cite{bhattacharya2012nonparametric} and \cite{tokdar2019bayesian}. In
both cases, the authors were able to prove the consistency under the
alternative hypothesis but not under the null. For a discussion on the
\virgolette{necessity} of these bounds in nonparametric models, see
\cite{tokdar2019bayesian}.

In light of the previous consistency result for the non-homogeneous case,
our recommendation to carry out the homogeneity test is to decide in favor
of $H_0$ whenever the posterior of $c_i, c_j$ does not strongly
concentrate on $c_i \neq c_j$. As Section~\ref{sec:experiments} shows, in
our simulated data experiments this choice consistently identifies the
right structure of homogeneity among populations. See also the discussion
later in Section~\ref{sec:disc}.

\section{Posterior Simulation}\label{sec:MCMC}

We illustrate an MCMC sampler based on the restaurant representation
derived in Section~\ref{sec:restaurant}. The random measures $\{F_i\}_i$
and $\Gtilde$ are marginalized out for all the updates except for the case
of $\bm c$, for which we use a result from \cite{pitman1996} to sample
from the full conditional of each $F_i$, truncating the infinite
stick-breaking sum adaptively; see below. We refer to this algorithm as
\emph{marginal}. 
We also note that, by a prior truncation of all the stick-breaking infinite sums to a fixed number of atoms,
we can derive a blocked Gibbs sampler as in
\cite{ishwaran2001gibbs}. However,  in our applications the
blocked Gibbs sampler was significantly slower both in reaching
convergence to the stationary distribution and to complete one single
iteration of the MCMC update. Hence, we will describe and use only the
marginal algorithm.

We follow the notation introduced in Section~\ref{sec:restaurant}. The
state of our MCMC sampler consists of the restaurant tables $\{
\theta^*_{rh} \}$, the tables in the common area $\{\tau_h\}$, a set of
binary variables $\{h_{rj}\}$, indicating if each table is
\virgolette{located} in the restaurant-specific or in the common area, a
set of discrete shared table allocation variables $t_{r\ell}$, one for
each $\theta^*_{r\ell}$ such that $\theta^*_{r\ell} = \tau_k$ iff
$t_{r\ell} = k$ and $h_{r\ell} = 0$,
the categorical variables $c_i$, indicating the restaurant for each
population, $\kappa \in (0, 1)$, and the table allocation variable
$s_{ij}$: for each observation such that $\theta_{ij} = \theta^*_{rh}$ iff
$c_i=r$ and $s_{ij} =h$. We also denote by $H_0$ and $H_r$ the number of
tables occupied in the shared area and in restaurant $r$ respectively,
$m_{rk}$ indicates the number of customers in the common area entered from
restaurant $r$ seating at table $k$.

We use the dot notation for marginal counts, for example $n_{r\bdot}$
indicates all the customers entered in restaurant $r$. We summarize
the Gibbs sampling scheme next.

\noindent
\begin{itemize}
\item Sample the cluster allocation variables using the Chinese
    Restaurant Process,
\begin{equation}\label{eq:cluster_alloc}
	p(s_{ij} = s \mid c_i=r, rest) \propto
		\begin{cases}
			n^{-ij}_{r\ell} k(y_{ij} \mid \theta^*_{rh}) \quad \text{if $s$  previously used} \\
			\alpha p(y_{ij} \mid \bm s^{-ij}, rest) \quad \text{if } s = s^{new}
			\end{cases}
	\end{equation}
where
\begin{multline}
\label{eq:marginal_clus_alloc}
	p(y_{ij} \mid \bm s^{-ij}, rest) = \kappa \int k(y_{ij} \mid \theta) G_0(d\theta) + \\
		+ (1 - \kappa) \left( \sum_{k=1}^{H_0} \frac{m^{-ij}_{\bdot k}}{m^{-ij}_{\bdot \bdot} + \gamma}
    k(y_{ij} \mid \tau_k) + \frac{\gamma}{m^{-ij}_{\bdot \bdot} + \gamma} \int k(y_{ij} \mid \theta) G_{00} (d\theta) \right),
\end{multline}
\noindent and where the notation $x^{-ij}$ means that
observation $y_{ij}$ is removed from the calculations involving the
variable $x$.

If $s=s^{new}$, a new table is created. The associated value
$\theta^*_{r s^{new}}$ is sampled from $G_0$ with probability $\kappa$
or from $\Gtilde$ with probability $1-\kappa$, as described in
Section~\ref{sec:restaurant}. The corresponding latent variables
$h_{rs^{new}}$ and $t_{rs^{new}}$ are set accordingly. When
sampling from \eqref{eq:crp_shared} a new table in the shared area
might be created. In that case, $t_{rs^{new}}$ is set to $H_0 + 1$.

\item Sample the table allocation variables $t_{r\ell}$ as in the HDP:
\begin{equation}\label{eq:table_alloc}
	p(t_{r\ell} = k \mid rest) \propto
		\begin{dcases}
		    m^{-r\ell}_{\cdot k} \prod_{(i, j): c_{i}=r, s_{i, j} = \ell} k(y_{ij} \mid \tau_k)  \quad \text{if $k$ previously used} \\
			\gamma \int \prod_{(i, j): c_{i}=r, s_{i, j} = \ell} k(y_{ij} \mid \tau) G_{00}(d\tau) \quad \text{if } k = k^{new},
			\end{dcases}
\end{equation}
where the notation $x^{-r\ell}$ means that table
$\theta^*_{r\ell}$, including all the associated observations, is
entirely removed from the calculations involving variable $x$. If
$k = k^{new}$ a new table is created in the shared area, the
allocation variables $s_{ij}$ are left unchanged.

\item Sample the cluster values from
	\begin{equation*}
		\Law(\theta^*_{r\ell} \mid h_{r\ell} = 1, rest) \propto G_0(\theta^*_{r\ell}) \prod_{(i, j): c_i = r, s_{i, j} =\ell} k(y_{ij} \mid \theta^*_{r\ell})
	\end{equation*}
	and
	\begin{equation*}
		\Law(\tau_{k} \mid rest) \propto G_{00} (\tau_k) \prod_{(i, j) \in (*)} k(y_{ij} \mid \tau_{k})
	\end{equation*}
	where the product $(*)$ is over all the index couples such that
$c_i=r, s_{ij} = \ell$, $h_{r\ell} = 0$ and $\theta^*_{r\ell} =
\tau_k$. Observe that, when $h_{r\ell} = 0$, it means that
    $\theta^*_{r\ell} = \tau_k$ for some $k$. Hence, in this case,
    $\theta^*_{r\ell}$ is purely symbolic and we do not need to sample
    a value for it.
\item Sample each $h_{r\ell}$ independently from
		\begin{align*}
			p(h_{r\ell} = 1|rest) &\propto \kappa G_{0}(\theta^*_{r\ell}) \\
			 p(h_{r\ell} = 0 |rest) &\propto (1-\kappa) \left(\sum_{k=1}^{H_0} \frac{m^{-r\ell}_{\bdot k}}{m^{-r\ell}_{\bdot \bdot} + \gamma}
\delta_{\tau_k}(\theta^*_{r\ell}) + \frac{\gamma}{m^{-r\ell}_{\bdot \bdot} + \gamma} G_{00} (\theta^*_{r\ell})\right),
			\end{align*} where, as in \eqref{eq:table_alloc}, the
notation $x^{-r\ell}$ means that table $\theta^*_{r\ell}$, including
all its associated observations, is removed from the calculations
involving variable $x$. Observe that, while in the update of the
cluster values all the $\theta^*_{r\ell}$ referring to the same
$\tau_k$ were updated at once, here we move the tables one by one.
\item Sample $\kappa$ from $\Law(\kappa \mid rest) \sim Beta
    \left(a_\kappa + \sum_{i,j} h_{ij}, \ b_\kappa + \sum_{i,j} (1 -
    h_{ij}) \right)$.
			
\item Sample $\bm \omega$ from

\[
    \bm \omega \mid rest \sim Dirichlet \Big(\eta_1 + \sum_{i=1}^I \mathbb{I}[c_i = 1], \ldots, \eta_I + \sum_{i=1}^I \mathbb I[c_i = I] \Big)
\]
where $\mathbb I[\cdot]$ denotes the indicator function.

\item Sample each $c_i$ in $\bm c = (c_1, \ldots, c_I)$ independently
    from
    \begin{equation}
		P(c_i = r \mid F_1, \ldots, F_I, \bm \omega, \bm y_i) \propto \omega_r \prod_{j=1}^{N_i} \int k(y_{ij} \mid \theta) F_r(d\theta) \label{eq:full_cond_c}.
    \end{equation}
If the new value of $c_i$ differs from the previous one, then
following \eqref{eq:cluster_alloc}, all the observations $y_{i1},
\ldots, y_{i N_i}$ are reallocated to the new restaurant.
\end{itemize}
Note that the update in \eqref{eq:full_cond_c} involves the previously
marginalized random probability measures $F_1, \ldots, F_I$. Thus, before
performing this update, we need to draw the $F_i$'s from their
corresponding full conditional distributions. It follows from Corollary 20
in \cite{pitman1996}  that the conditional distribution of $F_r$ given
$\bm c$, $\bm n_r$, $\bm \theta^*_r$, $\kappa$, and $\Gtilde$ coincides
with the distribution of $\pi_{r0} F^\prime_r + \sum_{h=1}^{H_r} \pi_{rh}
\delta_{\theta^*_{rh}}$, where $(\pi_{r0}, \pi_{r1}, \ldots, \pi_{rH_r})
\sim Dirichlet(\alpha, n_{r1}, \ldots, n_{rH_r})$ and $F^\prime_r \mid
\Gtilde \sim \mathcal{D}_{\alpha \Ptilde}$. This result was employed in
\cite{taddy2012mixture} to quantify posterior uncertainty of functionals
of a Dirichlet process, and also in \cite{canale2019importance} to derive
an alternative MCMC scheme for mixture models. It follows from the usual
stick breaking representation that $F^\prime_r = \sum_{h = 1}^\infty
w^\prime_{rh} \delta_{\theta^\prime_{rh}}$ with $\{ w^\prime_{rh}\}_h \sim
SB(\alpha)$ and $\theta^\prime_{rh} \mid \kappa, \Gtilde \iid \kappa G_0 +
(1-\kappa) \Gtilde$. Similarly, the conditional distribution of $\Gtilde$
given $\bm \tau$ and $\bm m$ coincides with the distribution of $v_0
\Gtilde^\prime + \sum_{k=1}^{H_0} v_k \delta_{\tau_k}$, where $(v_{0},
v_1, \ldots, v_{H_0}) \sim Dirichlet(\gamma, m_{\bdot 1}, \ldots, m_{\bdot
H_0})$ and $\Gtilde^\prime \sim \mathcal{D}_{\gamma G_{00}}$.

In practice, we draw each $F_r^\prime$ by truncating  the infinite sum.
Note that we do not need to set a priori the truncation level. Instead, we
can specify an upper bound for the error introduced by the truncation and
set the level adaptively. In fact, as a straightforward consequence of
Theorem 1 in \cite{ishwaran2002approximate}  we have that the total
variation distance between $F_r^\prime$ and its approximation with $M$
atoms, say $F_r^{M\prime}$, is bounded by $\varepsilon_M = 1 -
\sum_{h=1}^M w_{rh}^\prime$ \citep[see also Theorem 2
in][]{lijoi2020sampling}. The error induced on $F_r$ is then bounded by
$\pi_{r0} \varepsilon_M$. Note that simulation of the atoms
$\theta_{rh}^\prime$ involves the discrete measure $\Gtilde^\prime$.
However, we only need to draw a finite number of samples from it, and
not its full trajectory, so that no truncation is necessary for $\Gtilde^\prime$ . 
For ease of bookkeeping, we employ retrospective
sampling \citep{papaspiliopoulos2008retrospective} to simulate the atoms.
Alternatively, the classical CRP representation can be used. In our
experiments, because $\sum_{h=1}^{H_r} n_{rh} \gg \alpha$ we have
$\pi_{r0} \ll \sum_{h=1}^{H_r} \pi_{rh} \approx 1$. Thus, choosing a
truncation level $M=10$ always produces an error on $F_i$ lower than
$10^{-4}$ (henceforth fixed as the truncation error threshold).
Furthermore, we are often not even required to draw samples from $\Gtilde^\prime$.

Of the aforementioned steps, the bottleneck is the update of $\bm c$
because for each $c_i$ we are required to evaluate the densities of $N_i$
points in $I$ mixtures. If $N_i = N$ for all $i$, the computational cost
of this step is $O(NI^2)$, which can be extremely demanding for large
values of $I$. We can mitigate the computational burden by replacing this
Gibbs step with a Metropolis-within-Gibbs step, in the same spirit of the
\emph{Metropolised Carlin and Chib} algorithm proposed in
\cite{dellaportas2002}. At each step we propose a move from $c_i^{(\ell)}
= r$ to $c_i^{(\ell+1)} = m$ with a certain probability $p_i(m \mid r)$.
The transition is then accepted with the usual Metropolis-Hastings rule,
i.e. the new update becomes:
\begin{itemize}
\item Propose a candidate $m$ by sampling $p_i(m \mid r)$
\item Accept the move with probability $q$, where
		\[
			q = \min \left[1, \frac{P(c_i = m) \prod_{j=1}^{N_i} \int k(y_{ij} \mid \theta) F_m(d\theta)}{P(c_i = r)
\prod_{j=1}^{N_i} \int k(y_{ij} \mid \theta) F_r(d\theta)} \frac{p_i(r \mid m)}{p_i(m \mid r)} \right]
		\]
\end{itemize}
We call this alternative sampling scheme the Metropolised sampler. The key
point is that if evaluating the proposal $p_i(\cdot|\cdot)$ has a
negligible cost, the computational cost of this step will be $O(2NI)$ as
for each data point we need to evaluate only two mixtures: the one
corresponding to the current state $F_r$ and the one corresponding to the
proposed state $F_m$. Of course, the efficiency and mixing of the Markov
chain will depend on a suitable choice of the transition probabilities
$p_i(\cdot|\cdot)$; some possible alternatives are discussed in
Section~\ref{sec:experiments}.

When, at the end of an iteration, a cluster is left unallocated (or
empty), the probability of assigning an observation to that cluster will
be zero for all subsequent steps. As in standard literature, we employ a
relabeling step that gets rid of all the unused clusters. However, this
relabeling step is slightly more complicated since there are two different
types of clusters:  one arising from $G_0$ and ones arising from
$\Gtilde$. Details of the relabeling procedure are discussed in the
Appendix, Section~\ref{sec:s_relabeling}.

\subsection{Use of pseudopriors}\label{sec:pseudoprior}

The above mentioned sampling scheme presents a major issue that could
severely impact the mixing. Consider as an example the case when $I=2$;
if, at iteration $k$, the state jumps to $c_1 = c_2 = 1$, then all the
tables of the second restaurant would be erased from the state, because no
observation is assigned to them anymore. Switching back to $c_1 \neq c_2$
would then require that the approximation of $F_2$ sampled from its prior
distribution gives sufficiently high likelihood to either $\bm y_1$ or
$\bm y_2$, an extremely unlikely event in practice.

To overcome this issue, we make use of pseudopriors as in
\cite{carlin1995pseudoprior}, that is, whenever a random measure $F_r$ in
$(F_1, \dots, F_I)$ is not associated with any group, we sample the part
of the state corresponding to that measure (the atoms
$\{\theta^*_{r\ell}\}$ and number of customers $\{n_{r\ell}\}$ in each
restaurant) from its pseudoprior. From the computational point of view,
this is accomplished by running first a preliminary MCMC simulation where
the $c_i$'s are fixed as $c_i = i$, and collecting the samples.
Then, in the actual MCMC simulation, whenever restaurant $r$ is empty we
change the state by choosing at random one of the previous samples
obtained with fixed $c_i$'s. Note that this use of pseudopriors does not
alter the stationary distribution of the MCMC chain. Furthermore, the way
pseudopriors are collected and sampled from is completely arbitrary, and
our proposed solution works well in practice. Other valid options include
approximations based on preliminary chain runs, as discussed in
\cite{carlin1995pseudoprior}.

Section~\ref{sec:experiments} below contains extensive simulation studies
that show that the proposed model can be used to efficiently estimate
densities for each population. We also tried the case of a large number of
populations, e.g. $I=100$ without any significant loss of performance.

\section{Simulation Study}\label{sec:experiments}

In this section we investigate the ability of our model to estimate
dependent random densities. We fix the kernel $k(\cdot|\theta)$ in
\eqref{eq:latent_rep}  to be the univariate Gaussian density with
parameter $\theta = (\mu, \sigma^2)$ (mean and variance, respectively).
Both base measures $G_0$ and $G_{00}$ are chosen to be
$$
	\calN(\mu \mid 0, 10 \sigma^2)\times inv-gamma(\sigma^2 \mid 1, 1),
$$
and unless otherwise stated, with hyperparameters $\alpha, \gamma$ fixed
to 1, $a_\kappa = b_\kappa = 2$, and $\bm \eta = (1/I, \ldots, 1/I)$.
Chains were run for $100,000$ iterations after discarding the first
$10,000$ iterations as burn-in, keeping one every ten iterations,
resulting in a final sample size of $10,000$ MCMC draws.

\subsection{Two populations}

We first focus on the special case of $I=2$ populations. Consider
generating data as follows
\begin{equation}
\begin{aligned}
	y_{1j} &\iid w_1 \calN(\mu_1, \sigma_1) + (1 - w_1) \calN(\mu_2, \sigma_2) \quad j=1, \dots N_1\\
	y_{2j} &\iid w_2 \calN(\mu_3, \sigma_3) + (1 - w_2) \calN(\mu_4, \sigma_4) \quad j=1, \dots N_2,
\end{aligned}
\label{eq:two_pop_data}
\end{equation}
that is each population is a mixture of two normal components. This is the
same example considered in \cite{Cam_etal_2018latent}.
\begin{table}
\centering
\begin{tabular}{ |c||c|c|c|c|c|c| }
 \hline
 &$(\mu_1, \sigma_1)$ & $(\mu_2, \sigma_2)$ & $(\mu_3, \sigma_3)$ & $(\mu_4, \sigma_4)$ & $w_1$ & $w_2$  \\
 \hline
 Scenario I & (0.0, 1.0) & (5.0, 1.0) & (0.0, 1.0) & (5.0, 1.0) & 0.5 & 0.5 \\
 \hline
 Scenario II & (5.0, 0.6) & (10.0, 0.6) & (5.0, 0.6) & (0.0, 0.6) & 0.9 & 0.1 \\
 \hline
 Scenario III & (0.0, 1.0) & (5.0, 1.0) & (0.0, 1.0) & (5.0, 1.0) & 0.8 & 0.2 \\
 \hline
\end{tabular}
\caption{Parameters of the simulated datasets}
\label{tab:simulated_params}
\end{table}
Table~\ref{tab:simulated_params} summarizes the parameters used to
generate the data. Note that these three scenarios cover either the full
exchangeability case across both populations (Scenario I), as well as the
partial exchangeability between the two populations (scenarios II and
III). For each case, we simulated $N_1=N_2=100$ observations for each
group (independently).

\begin{figure}[t]
\centering
\begin{subfigure}{\linewidth}
\centering
	\includegraphics[width=0.6\linewidth]{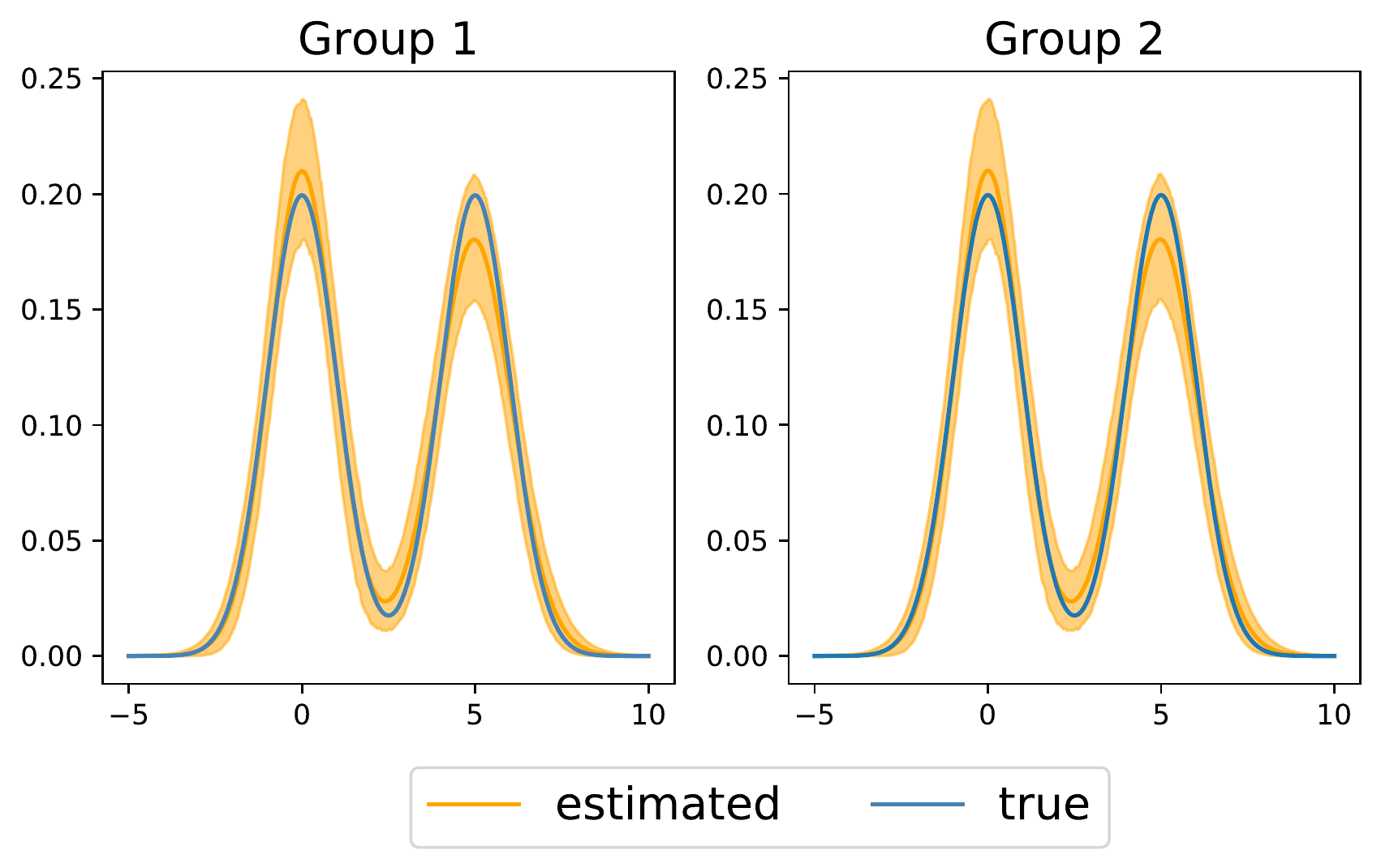}
\end{subfigure}
\begin{subfigure}{\linewidth}
\centering
	\includegraphics[width=0.6\linewidth]{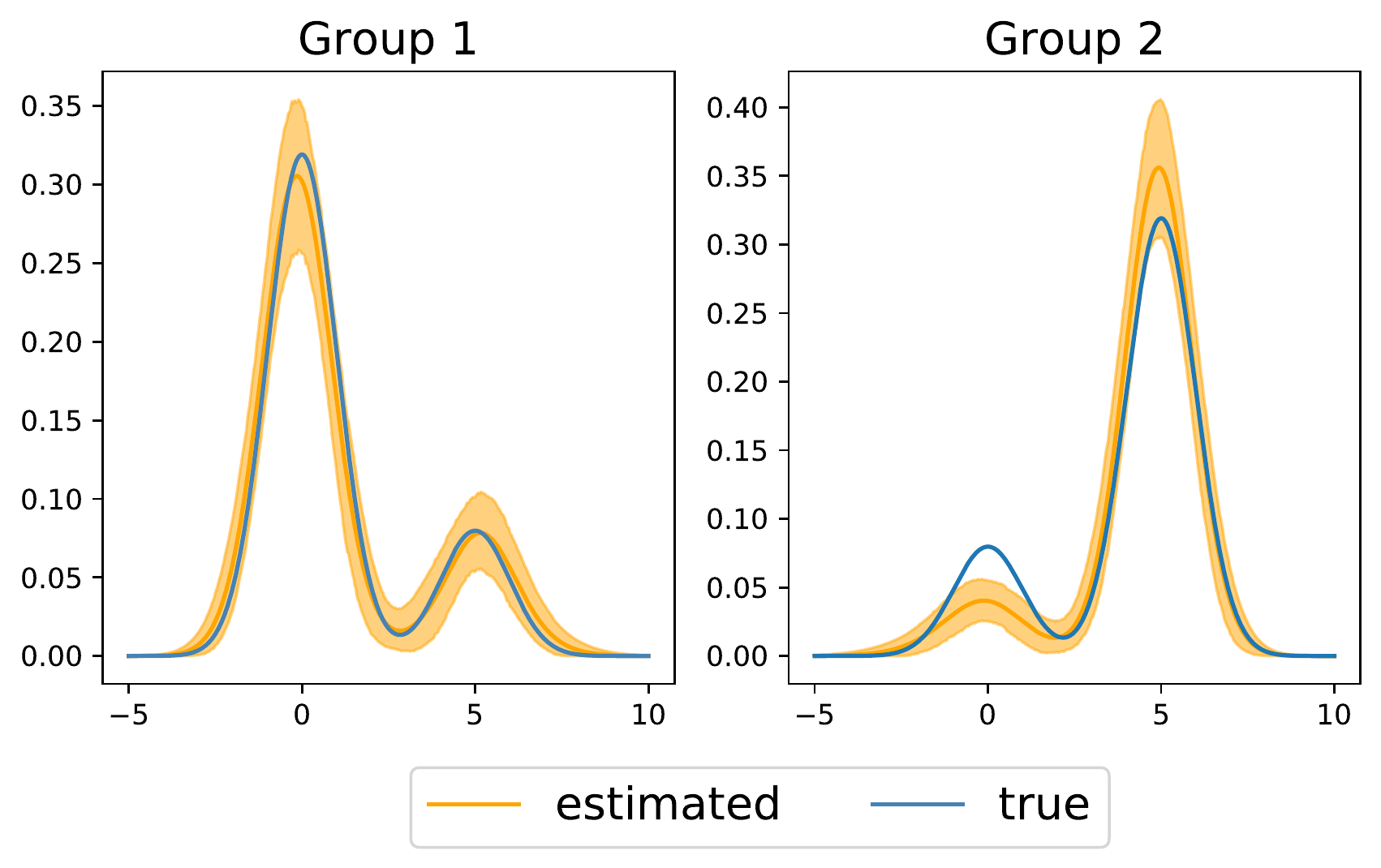}
\end{subfigure}
\caption{Density estimates and pointwise $95\%$ posterior
credible intervals for the two populations of Scenario I (top) and Scenario
III (bottom).}
\label{fig:two_pop_density}
\end{figure}

Table~\ref{tab:two_pop_results} reports the posterior probabilities of the
two population being identified as equal for the three scenarios. We can
see that our model recovers the ground truth. Moreover
Figure~\ref{fig:two_pop_density} shows
the density estimates, i.e. the posterior mean of the density evaluated on
a fixed grid of points, together with pointwise $95\%$ posterior credible
intervals at each point $x$ in the grid, obtained by our MCMC for
scenarios I and III. Here, densities are estimated from the
corresponding posterior mean evaluated on a fixed grid of points, while
credible intervals are obtained by approximating the $F_i$'s as discussed
in Section~\ref{sec:MCMC}. We can see that in both the cases,
locations and scales of the populations are recovered perfectly, while it
seems that the weights of the mixture components are slightly more precise
in Scenario I than in Scenario III.

\begin{table}[h]
\centering
\begin{tabular}{ |c|c|c| }
 \hline
  & $P(c_1 = c_2 \mid data)$ & $BF_{01}$  \\
  \hline
  Scenario I & 0.99 & 98.9  \\
  Scenario II & 0.0 & 0.0 \\
  Scenario III & 0.0 & 0.0 \\
  \hline
\end{tabular}
\caption{Posterior inference}
\label{tab:two_pop_results}
\end{table}

\begin{figure}[t]
\centering
\includegraphics[width=0.9\linewidth]{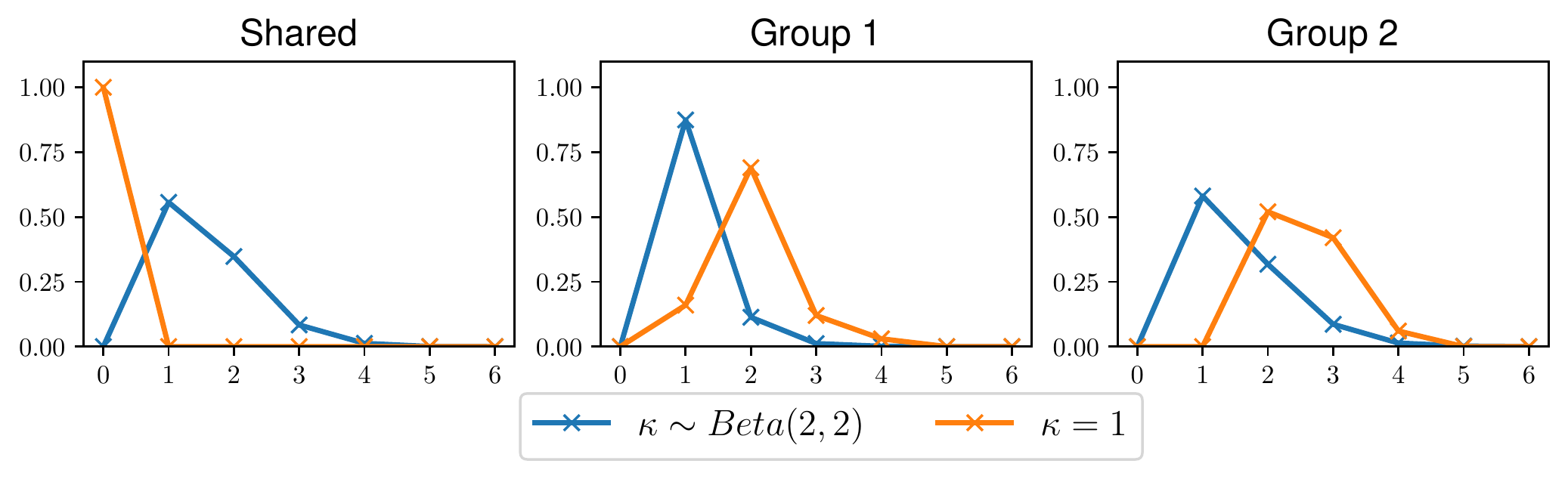}
\caption{Posterior distribution of the number of shared unique values and unique
values specific to first and second group in Scenario II.}
\label{fig:cluster_distrib}
\end{figure}

Comparing the Bayes Factors shown in Table~\ref{tab:two_pop_results} with
the ones in \cite{Cam_etal_2018latent} (5.86, 0.0 and 0.54 for the three
scenarios, respectively), we see that both models are able to correctly
assess homogeneity. However, the Bayes Factors obtained under our model
tend to assume more extreme than those from~\cite{Cam_etal_2018latent}.
Figure~\ref{fig:cluster_distrib} shows the posterior distribution of the
number of shared and private unique values (reconstructed from the cluster
allocation variables $s_{ij}$ and the table allocation variables
$t_{r\ell}$) in Scenario II, when either $\kappa \sim Beta(2, 2)$ or
$\kappa = 1$. Also in the he latter case $P(c_1 = c_2 \mid  data) = 0$,
but the shared component between groups one and two is not recovered, due
to the degeneracy issue described in Proposition~\ref{prop:degeneracy}.

\begin{figure}[t]
\centering
\includegraphics[scale=0.8]{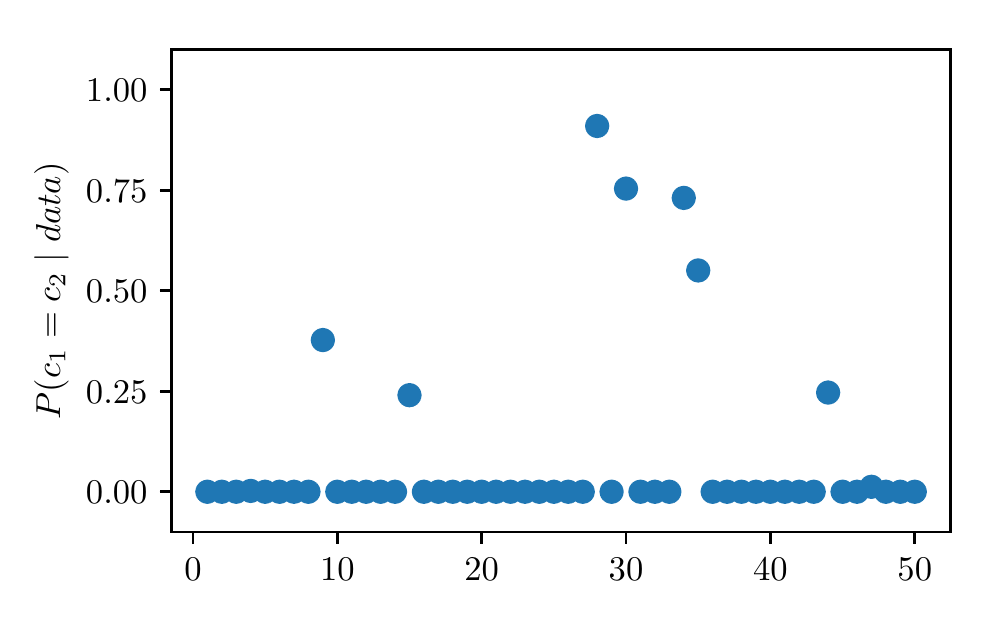}
\caption{Plot of the posterior probabilities $P(c_1 = c_2|data)$ for all of the 50 simulated datasets.
} \label{fig:two_population_meta}
\end{figure}

As the central point of our model is to allow for different random
measures to share at least one atom,
we test more in detail this scenario. To do so, we simulate 50 different
datasets from \eqref{eq:two_pop_data}, by selecting $\mu_1, \mu_2, \mu_4
\iid \calN(0, 10)$ and $\sigma^2_1, \sigma^2_2, \sigma^2_4 \iid
inv-gamma(2, 2)$, $w_1 \sim Beta(1, 1)$ and setting $\mu_3 = \mu_1,
\sigma^2_3 = \sigma^2_1, w_2 = w_1$. In this way we create 50 independent
scenarios where the two population share exactly one component and give
the same weight to this component. Figure~\ref{fig:two_population_meta} reports the scatter plot of the estimated
posterior probabilities of $c_1 = c_2$ obtained from the MCMC samples. It
is clear that our model recovers the right scenario most of the times. Out
of 50 examples, only in four of them $P(c_1 = c_2 \mid data)$ is greater
than 0.5, by a visual analysis we see from the plot of the true densities
that in those cases the two populations were really similar.

\subsection{More than two populations}
\label{sec:morethan2}

We extend now the simulation study to scenarios with more than two
populations. We consider three simulated datasets with four populations
each and different clustering structures at the population level. In
particular, we use the same scenarios as in \cite{gutierrez2019}, and
simulate $N_i = 100$ points for each population $i = 1, 2, 3, 4$ as
follows
\begin{itemize}
\item Scenario IV
	\[
	y_{1j}, y_{2k}, y_{3\ell} \iid \calN(0, 1) \quad y_{4n} \iid SN(0, 1, 1) \quad j, k, \ell, n = 1, \ldots, 100
	\]
\item Scenario V
	\[
	y_{1j}, y_{4n} \iid \calN(0, 1) \quad y_{2k} \iid \calN(0, 2.25) \quad y_{3\ell} \iid \calN(0, 0.25) \quad j, k, \ell, n = 1, \ldots, 100
	\]

\item Scenario VI
	\begin{align*}
		y_{1j}, y_{2k} &\iid 0.5 \calN(0, 1) + 0.5 \calN(5, 1)  \quad j, k = 1, \ldots, 100\\
		y_{3\ell} &\iid 0.5\calN(0, 1) + 0.5\calN(-5, 1)  \quad \ell = 1, \ldots, 100 \\
		y_{4n} &\iid 0.5\calN(-5, 1) + 0.5\calN(5, 1) \quad n = 1, \ldots, 100
	\end{align*}
\end{itemize}
Hence, the true clusters of the label set of the populations, $\{1, 2, 3,
4\}$, are: $\bm \rho^{true}_4 = \{\{1, 2, 3\}, \{4\}\}$, $\bm
\rho^{true}_5 = \{\{1, 4\}, \{2\}, \{3\}\}$ and $\bm \rho^{true}_6 =
\{\{1, 2 \}, \{3\},  \{4\}\}$ for the three scenarios under investigation
respectively. By $SN(\xi, \omega, \alpha)$ in Scenario IV we mean the
skew-normal distribution with location $\xi$, scale $\omega$ and shape
$\alpha$; in this case, the mean of the distribution is equal to
\[
    \xi + \omega \frac{\alpha}{1+\alpha^2}\sqrt{\frac{2}{\pi}} .
\]

Note that we focus on a different problem than what \cite{gutierrez2019}
discussed, as they considered testing for multiple treatments against a
control. In particular they were concerned about testing the hypothesis of
equality in distribution between data coming from different treatments
$\bm y_{j}$ ($j=2, 3, 4$ in these scenarios), and data coming from a
control group $\bm y_{1}$. Instead our goal is to cluster these
populations based on their distributions.

Observe how the prior chosen for $\bm c$ does not translate directly into
a distribution on the partition $\bm \rho$, as it is affected by the so
called label switching. Thus, in order to summarize our inference, we
post-process our chains and transform the samples  $\bm c^{(1)}, \ldots,
\bm c^{(M)}$ from $\bm c$ to samples $\bm \rho^{(1)},\ldots,\bm
\rho^{(M)}$ from $\bm \rho$. For example we have that $\bm c^{(i)} = (1,
1, 1, 3)$ and $\bm c^{(j)} = (2, 2, 2, 4)$ both get transformed into $\bm
\rho^{(i)} = \bm \rho^{(j)} = \{\{1, 2, 3\}, \{4\}\}$.

\begin{figure}[t]
    \centering
		\includegraphics[width=0.5\linewidth]{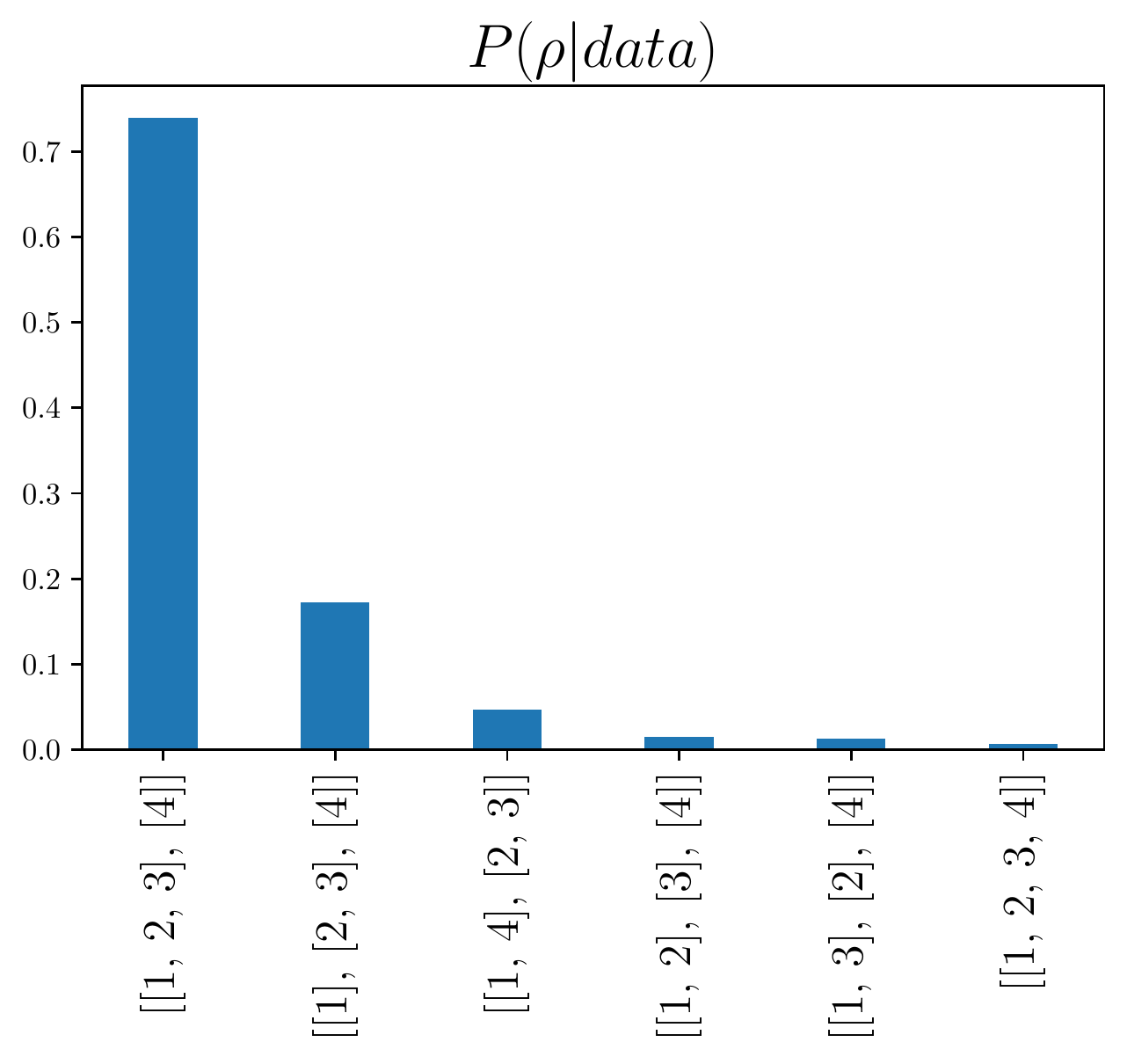}
	\caption{Posterior probability of $\bm \rho$ for Scenario IV.}
	\label{fig:four_pop_exp1}
\end{figure}

\begin{figure}[t]
\centering
\includegraphics[width=0.9\linewidth]{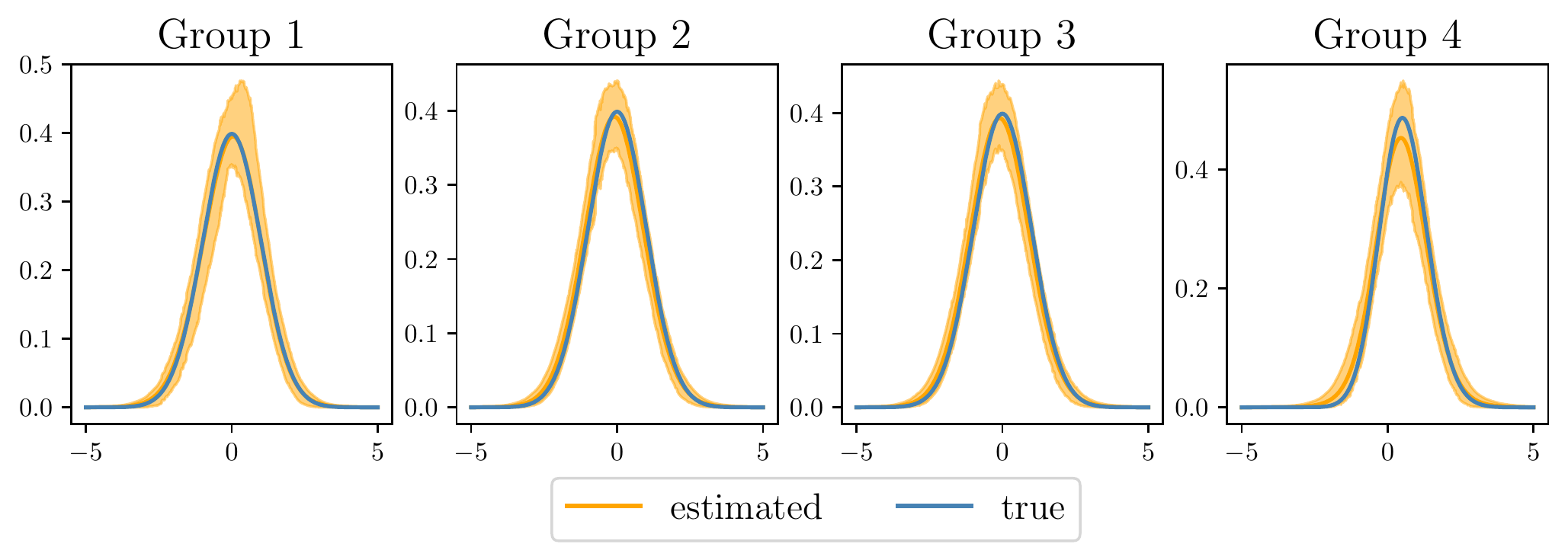}
\caption{Density estimates and pointwise 95\% posterior credible intervals for
Scenario IV.}
\label{fig:scen4_dens}
\end{figure}

The posterior probabilities of the true clusters $P(\bm \rho_i = \bm
\rho_i^{true} \mid \rm{data})$ are estimated using the transformed (as
described above) MCMC samples and equal 0.75, 0.95 and 0.99 for the three
scenarios respectively. Figure~\ref{fig:four_pop_exp1} shows the posterior
distribution of $\bm \rho$, and Figure~\ref{fig:scen4_dens} reports the
density estimation of each group, for Scenario IV. Observe how the
posterior mode is in $\bm \rho_4^{true}$ but significant mass is given
also to the case $\{\{1\}, \{2, 3\}, \{4\}\}$. We believe that this
behavior is mainly due to our use of pseudopriors, as it makes the
transition between these three states fairly smooth. On the other hand, in
Scenario V, where the posterior mass on the true cluster is close to 1, it
is clear that such transitions happen very rarely, as the posterior
distribution, not shown here, is completely concentrated on $\bm
\rho_5^{true}$. Our insight is that the pseudopriors make a transition
between two states, say $\bm c^{(j)} = (1, 1, 3, 4)$ and $\bm c^{(j+1)} =
(1, 2, 3, 4)$ (or viceversa), more likely when the mixing distributions of
population one and two are the same.

We compared the performance of the Metropolised algorithm against the full
Gibbs move for the update of $\bm c$, computing the effective sample size
(ESS) of the number of population level clusters (i.e. the number of
unique values in $\bm c$) over CPU time. We consider two choices for the
proposal distribution $p_i(r \mid m)$, namely, the discrete uniform over
$\{1, \ldots, I\}$ and another discrete alternative, with weights given by
\begin{equation}
	p_i(r \mid m) \propto 1 + \left(1 + d^2(F_r, F_m) \right)^{-1}
	\label{eq:metro_proposal}
\end{equation}
where $d^2(F_r, F_m)$ is the squared $L^2$ distance between the Gaussian
mixture represented by $F_r$ and that represented by $F_m$, which are
sampled as discussed in Section~\ref{sec:MCMC}.
Let $p_r = \sum_{i=1}^{H_r} w_{ri} \calN(\mu_{ri}, \sigma^2_{ri})$ and $p_m =
\sum_{j=1}^{H_m} w_{mj} \calN(\mu_{mj}, \sigma^2_{mj})$ be the mixture densities associated to the mixing measures $F_r$ and $F_m$ respectively, then:
\begin{equation}\label{eq:mix_dist}
\begin{aligned}
	d^2(F_r, F_m) &= L_2(p_r, p_m)^2 = \sum_{i, i^\prime =1}^{H_r} w_{ri}, w_{ri^\prime} \int \calN(y;\mu_{ri}, \sigma^2_{ri}) \calN(y;\mu_{ri^\prime}, \sigma^2_{ri^\prime}) dy  \\
	& \qquad\qquad \qquad
	+\sum_{j, j^\prime=1}^{H_m} w_{mj}, w_{mj^\prime} \int \calN(y;\mu_{mj}, \sigma^2_{mj}) \calN(y;\mu_{mj^\prime}, \sigma^2_{mj^\prime}) dy
	\\
	& \qquad\qquad \qquad -2 \sum_{i=1}^{H_r} \sum_{j=1}^{H_m} w_{ri} w_{mj} \int \calN(y; \mu_{ri}, \sigma^2_{ri}) \calN(y;\mu_{mj}, \sigma^2_{mj}) dy.
\end{aligned}
\end{equation}
which can be easily computed in closed form since
\begin{equation}\label{eq:normal_prod_int}
\int \calN(y;\mu, \sigma^2) \calN(y;\mu^\prime, (\sigma^{\prime})^2) dy = \calN(\mu; \mu^\prime, \sigma^2 + (\sigma^{\prime})^2).
\end{equation}
See the Appendix, Section~\ref{sec:s_proofs}, for the proof of Equations \eqref{eq:mix_dist}-\eqref{eq:normal_prod_int}.

Results for data as in Scenario IV show that the best efficiency is
obtained using the full Gibbs update, with an ESS per second of 57.1. The
Metropolised sampler with proposal as in \eqref{eq:metro_proposal} comes
second, yielding an ESS per second of 34.1 while the Metropolised sampler
with uniform proposal is the worst performer with an ESS per second of
12.8. Hence, even when the number of groups is not enormous, the good
performance of the Metropolised sampler is clear. Preliminary analysis
showed how the Metropolised sampler outperforms the full Gibbs one as the
number of groups increases.

Finally, we test how our algorithm performs when the number of populations
increases significantly. We do so by generating $100$ populations in
Scenario VII as follows:
\begin{align*}
	y_{ij} &\iid 0.5 \calN(-5, 1) + 0.5 \calN(5, 1) \quad i = 1, \ldots, 20\\
	y_{ij} &\iid 0.5 \calN(-5, 1) + 0.5 \calN(0, 1) \quad i = 21, \ldots, 40\\
	y_{ij} &\iid 0.5 \calN(0, 1) + 0.5 \calN(5, 0.1) \quad i = 41, \ldots, 60\\
	y_{ij} &\iid 0.5 \calN(-10, 1) + 0.5 \calN(0, 1) \quad i = 61, \ldots, 80\\
	y_{ij} &\iid 0.1 \calN(-10, 1) + 0.9 \calN(0, 1) \quad i = 81, \ldots, 100.
\end{align*}
Thus, full exchangeability holds within populations $\{1, \ldots, 20\}$,
$\{21, \ldots, 40\}$, $\{41, \ldots, 60\}$, $\{61, \ldots, 80\}$ and
$\{81, \ldots, 100\}$ but not between these five groups.
For each population $i$, $100$ datapoints were sampled independently.

To compute posterior inference, we run the Metropolised sampler with
proposal \eqref{eq:metro_proposal}. To get a rough idea of the
computational costs associated to this large simulated dataset, we report
that running the full Gibbs sampler would have required more than 24 hours
on a 32-core machine (having parallelized all the computations which can
be safely parallelized), while the Metropolised sampler ran in less than 3
hours on a 6-core laptop.

As a summary of the posterior distribution of the random partition $\bm
\rho_{100}$, we compute the posterior similarity matrix $[P(c_i = c_j \mid
data)]_{i,j = 1}^I$.
Estimates of these probabilities are straightforward to obtain using the
output of the MCMC algorithm. Figure~\ref{fig:100_pop} shows the posterior
similarity matrix as well as the density estimates of five different
populations. It is clear that the clustering structure of the populations
is recovered perfectly and that the density estimates are coherent with
the true ones.

\begin{figure}[ht]
\centering
\includegraphics[width=\linewidth]{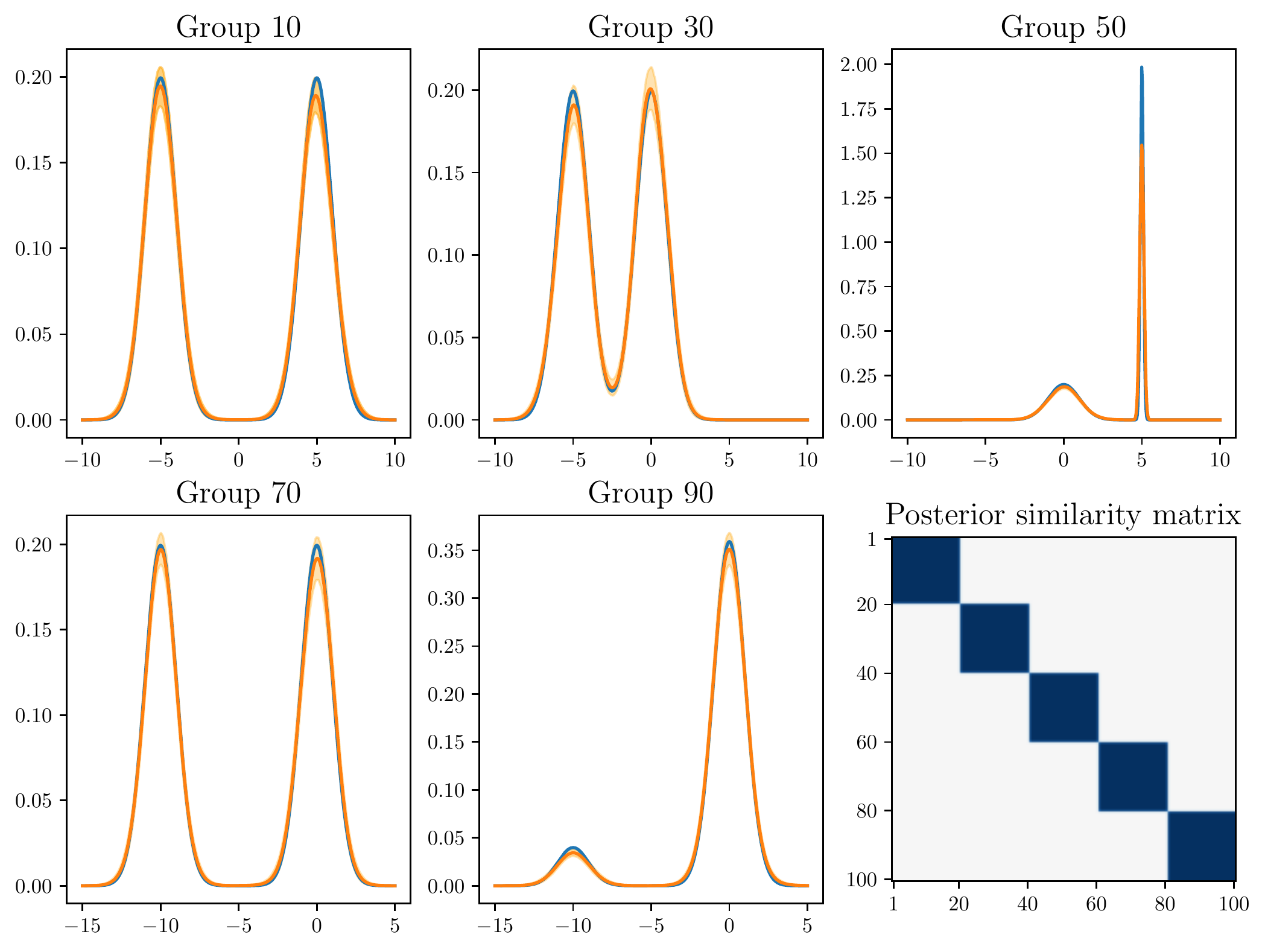}
\label{fig:100_pop}
\caption{Density estimates (orange line), pointwise 95\% posterior credible
intervals (orange bands), true data generating densities (blue line) for
groups 10, 30, 50, 70 and 90 and posterior similarity matrix (bottom
right, white corresponds to 0.0 and dark blue to 1.0) in Scenario VII.}
\label{fig:100_pop}
\end{figure}

\subsection{A note on the mixing}

One aspect of the inference presented so far that is clear from all the
simulated scenarios, is that the posterior simulation of $\bm c$, and
hence of the partition $\bm \rho$, usually stabilizes around one
particular value and then very rarely moves. This could be interpreted as
a mixing issue of the MCMC chain. However, notice that once the
\virgolette{true} partition of the population is identified, it is
extremely unlikely to move from that state, which can be seen directly
from Equation \eqref{eq:full_cond_c}. Indeed, moving from one state to
another modifies the likelihood of an entire population. In particular,
moving from a state where $c_i = c_j$ for two populations $i$ and $j$ that
are \emph{actually} homogeneous, to a state where $c_i \neq c_j$ is an
extremely unlikely move.

To further illustrate the point, consider for ease of explanation the case
of Simulation Scenario I where both populations are the same, and suppose
that at a certain MCMC iteration we impute $c_1 = c_2 = 1$. In order for
the chain to jump to $c_2 = 2$, the \virgolette{empty} mixing distribution
$F_2$ must be sampled in such a way to give a reasonably high likelihood
to all the data from the second population $y_{21}, \ldots y_{2N_2}$;
again, see \eqref{eq:full_cond_c}. If one did not make use of
pseudopriors, this would mean that $F_2$ would be sampled from the prior,
thus making this transition virtually impossible. But even using
pseudopriors, the transition remains quite unlikely. Indeed, once $c_1 =
c_2 = 1$, we get an estimate of $F_1$ using data from the two homogeneous
groups, hence getting a much better estimate that one would get when $c_1
\neq c_2$.

Nevertheless, in all simulation scenarios we tried this problem has not
prevented the posterior simulation algorithm from identifying the correct
partition of populations, as defined in these scenarios. In particular, we
found that $P(\bm \rho_4^{true}|data) = 0.75$  only in scenario IV , while
in all the other cases we tried,  the values of $P(\bm
\rho_4^{true}|data)$ was greater than 0.9. We also computed the cluster
estimate of the posterior of $\bm\rho$ that minimizes the posterior
expectation of Binder's loss  \citep{Binder78} under equal
misclassification costs and of the variation of information loss
\citep{wade2018bayesian}. In all the examples proposed, the
\virgolette{true} partition was correctly detected by both estimates.

\section{Chilean grades dataset}\label{sec:data}

The School of Mathematics at Pontificia Universidad Cat\'olica de Chile
teaches many undergraduate courses to students from virtually all fields.
When the number of students exceeds a certain maximum pre-established
quota, several sections are formed, and courses are taught in parallel.
There is a high degree of preparation in such cases, so as to guarantee
that courses cover the same material and are coordinated to function as
virtual copies of each other. In such cases, only the instructor changes
across sections, but all materials related to the courses are the same,
including exams, homework, assignments, projects, etc., and there is a
shared team of graders that are common to all the parallel sections.
According to the rules, every student gets a final grade on a scale from
1.0 to 7.0, using one decimal place, where 4.0 is the minimum passing
grade. We consider here the specific case of a version of Calculus II,
taught in parallel to three different sections (A, B and C) in a recent
semester. Our main goal here is to examine the instructor effectiveness,
by comparing  the distributions of the final grades obtained by each of
the three populations (sections). The sample sizes of these populations
are 76, 65 and 50 respectively.

A possible way to model these data could be to employ a truncated normal
distribution as the kernel in \eqref{eq:latent_rep}. However since our
primary interest is to investigate the homogeneity of the underlying
distributions and not to perform density estimates, we decided to first
add a small amount of zero-mean Gaussian noise, with variance $0.1$ to the
data (i.e. \virgolette{jittering}) and then   proceeded to standardize the
whole dataset, by letting $y_{ij}^{new} = (y_{ij} - \bar{y}) / s_y$, where
$\bar{y} = (\sum_{ij}y_{ij})/(\sum_i N_i)$ and $s^2_y = (\sum_{ij} (y_{ij}
- \bar{y})^2)/(\sum_i N_i -1)$ are the global sample mean and variance,
respectively. In the sequel, index $i=1,2,3$ denotes sections A, B and C,
respectively, as described above.

\begin{figure}[t]
\centering
\begin{subfigure}[t]{0.75\linewidth}
\vskip 0pt
\includegraphics[scale=0.8]{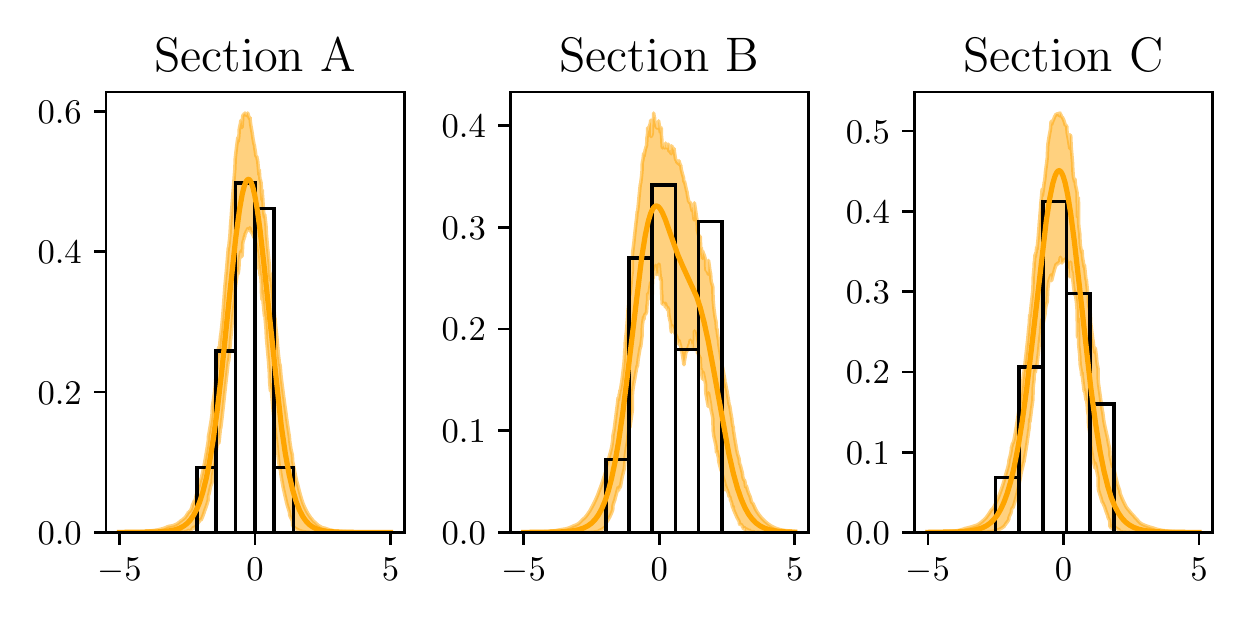}
\end{subfigure}%
\begin{subfigure}[t]{0.25\linewidth}
	\vskip 0pt
\includegraphics[width=\linewidth]{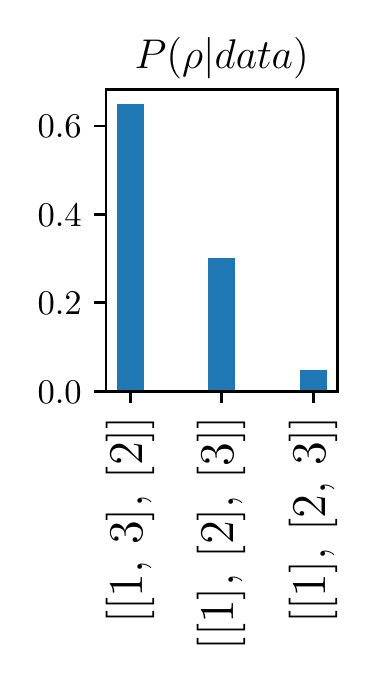}
\end{subfigure}
\vspace{-15pt}
\caption{Density estimates and pointwise 95\% posterior credible intervals for the
three groups (left); posterior distribution of the clusters (right).}
\label{fig:chile_plots}
\end{figure}

Figure~\ref{fig:chile_plots} reports density estimates in all groups (i.e.
posterior density means evaluated on a fixed grid of points and pointwise
$95\%$ posterior credible intervals at each point $x$ in the grid), as
well as  the posterior distribution of the random partition $\bm \rho$,
obtained from the posterior distribution of $\bm c$, getting rid of the
label switching in a post-processing step (see also
Section~\ref{sec:morethan2}). From Figure~\ref{fig:chile_plots} we see
that the posterior distribution of $\bm\rho$ gives high probability to the
case of the three groups being all different as well as to the case when
the first and third groups are homogeneous but different from the second
one. This is in accordance with a visual analysis of the observed and
estimated densities.

We considered several functionals of the random population distribution
$F_{c_i}$ (see \eqref{eq:hdp_lik}) for $i=1,2,3$.  Recall that, according
to notation in \eqref{eq:mixtHi}, $F_{c_i}=G_i$. First of all, we consider
the mean and variance functionals of the random density
$p_i(y)=\int_\Theta k(y|\theta) F_{c_i}(d\theta)=\int_\Theta k(y|\theta)
G_i(d\theta)$, for each $i=1,2,3$. Observe how they are functionals of the
random probability $F_{c_i}=G_i$. Moreover, since
Figure~\ref{fig:chile_plots}  seems to suggest that the three groups
differ mainly due to their different asymmetries, we considered two more
functionals of $G_i$, i.e. two indicators of skewness: Pearson's moment
coefficient of skewness $\text{sk}$ and the measure of skewness with
respect to the mode $\gamma_M$ proposed by \cite{arnold1995measuring}.
Pearson's moment coefficient of skewness of the random variable $T$ is
defined as $\text{sk} = \E [( (T - \E(T)) / \sqrt{\Var(T)})^3 ] $, while
the measure of skewness with respect to the mode
as $\gamma_M = 1 - 2 F_T(M_T)$, where $M_T$ is the mode of $T$ and $F_T$
denotes its distribution function. The last functional of $G_i$ we
consider is the probability, under the density $p_i(y)=\int_\Theta
k(y|\theta) G_i(d\theta)$ of getting a passing grade ($\geq 4.0$ before
normalization), that is
\[P_{4i}= \int_{(4-\bar y)/s_y}^{+\infty} p_i(y) dy . \]

Table~\ref{tab:chil_summary} shows  the posterior mean of the functionals
$\mu_i$, $\sigma_i^2$ (mean and variance functionals), $sk_i$,
$\gamma_{Mi}$ and $P_{4i}$ of $p_i$, for $i=1,2,3$.
\begin{table}[t]
\centering
\begin{tabular}{ |c|c|c|c|c|c| }
 \hline
  Section & ${\mu_i}$ & $ \sigma_i^2$ & $ {\text{sk}_i}$  & $ {\gamma_{Mi}}$ & $ {P_{4i}}$ \\
  \hline
   A & -0.264 & 0.671 & 120.84 & -0.01 & 0.53\\
   B & 0.438 & 1.428 & -64.86 & 0.292 & 0.71 \\
   C & -0.171 & 0.943 & 55.60 & -0.01 & 0.56 \\
  \hline
\end{tabular}
\caption{Posterior means of functionals $\mu_i,\ldots, P_{4i}$ of the
population density $p_i$ for each Section A ($i=1$), B ($i=2$) and C
($i=3$) in the Chilean grades dataset. All the functionals refer to
standardized data $\{y_{ij}^{new}= (y_{ij} - \bar{y})/ s_y \}$.  }
 \label{tab:chil_summary}
\end{table}
To be clear, the posterior mean of the mean functional $\mu_1$ is computed as
\[ \frac{1}{M} \sum_{\ell=1}^M \mu_1^{(\ell)} = \frac{1}{M} \sum_{\ell=1}^M \E[y \mid G_1^{(\ell)}] = \frac{1}{M} \sum_{\ell=1}^M  \left( \int_\Rea y p_1^{(\ell)}(y) dy\right),
\]
where $M$ is the MCMC sample size, and the superscript $(\ell)$ attached
to a random variable denotes its value at the $\ell$--th MCMC iteration.

In agreement with the posterior distribution of the partition $\bm \rho$,
for all the  functionals considered we observed close values for sections
A and C, while both differ significantly from the values for section B. In
summary, we conclude that section B presents a heavier right tail than
sections A and C, hence it is characterized by a higher mean (positive)
and also more spread across the range. Section B shows a larger
(estimated) value for $P_4$, i.e. students in section B are more likely to
pass the exam than their colleagues from the other sections. This seems to
suggest that a higher concentration of good students (with high grades)
was present in Section B, compared to A and C, possibly combined with a
higher effectiveness of the instructor in this Section.

\begin{figure}[t]
    \centering
    \includegraphics[width=\linewidth]{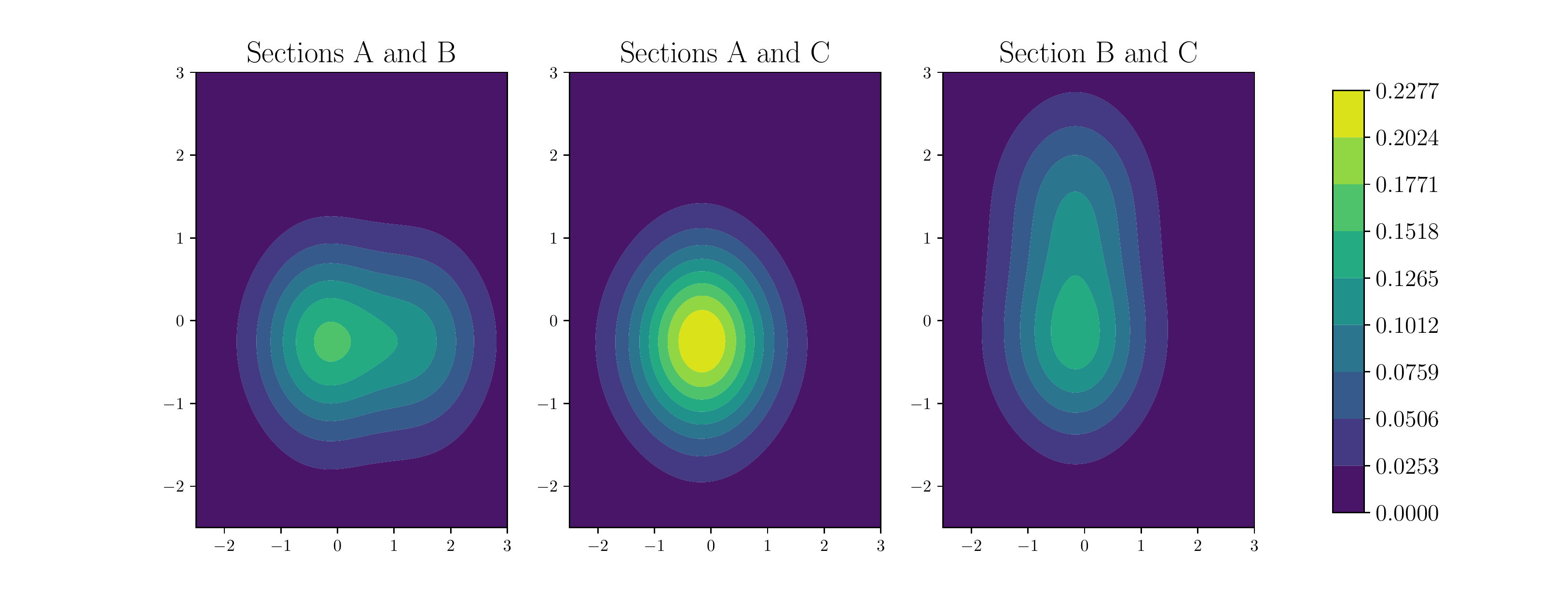}
    \caption{Posterior means of $(p_i,p_{\ell})$, $i \neq \ell$, $i,\ell=A,B,C$,
evaluated on a fixed grid in $\Rea^2$ for the Chilean grades dataset.}
    \label{fig:chil_joint}
\end{figure}

We also computed the pairwise $L^1$ distances between the estimated
densities in the  populations. If $\tilde p_i$ denotes the estimated
density (posterior mean of $p_i$ evaluated in a grid of points) for each
population, we found  $d(\tilde p_A, \tilde p_B) = 0.56$, $d(\tilde p_A,
\tilde p_C) = 0.15$ and $d(\tilde p_B, \tilde p_C) = 0.44$. This confirms
once again that the estimated densities for section A and C are closer
than when comparing sections A and B and sections B and C.

To end the analysis, we show in Figure~\ref{fig:chil_joint} estimated
couples of densities $(p_i,p_{\ell})$, $i\neq \ell$, $i,\ell=1,2,3$, i.e.
the posterior mean of $(p_i,p_{\ell})$, evaluated on a fixed grid in
$\Rea^2$. While sections A and C look independent (central panel in
Figure~\ref{fig:chil_joint}), the (posterior) propensity of section B to
get higher grades is confirmed in the left and right panels in
Figure~\ref{fig:chil_joint}.

\section{Discussion}\label{sec:disc}

Motivated by the traditional problem of testing homogeneity across $I$
different groups or populations, we have presented a model that is able to
not only address the problem but also to perform a cluster analysis of the
groups. The model is built on a prior for the population distributions
that we termed the semi-hierarchical Dirichlet process, and it was shown
to have good properties and also to perform well in synthetic and real
data examples, also in case of $I=100$ groups. One of the driving features
of our proposal was to solve the degeneracy limitation of nested
constructions that has been pointed out by~\cite{Cam_etal_2018latent}. The
crucial aspect of the semi-HDP that solves this problem was described
using the metaphor of a food court of Chinese restaurants with common and
private dining area. The hierarchical construction introduces a random
partition at the population level, which allows for identifying possible
clusters of internally homogeneous groups.

Our examples focus on unidimensional data, though extensions to
multivariate responses can be straightforwardly accommodated in our
framework. However, scaling with respect to data dimension is not a
property we claim to have. In fact, this is a situation shared with any
type of hierarchical mixture models.

We studied support properties of the semi-HDP and also the posterior
asymptotic behavior of the Bayes factor for the  homogeneity test when
$I=2$, as posed within the proposed hierarchical construction. We showed
that the Bayes factor has the appropriate asymptotic behavior under the
alternative hypothesis of partial exchangeability,  but a final answer
under the assumption of truly exchangeable data is still pending. The lack
of asymptotic guarantees is not at all specific to our case.  In fact,
this situation is rather common to all model selection problems when the
hypothesis are not well separated and at least one of the two models under
comparison is \virgolette{truly} nonparametric, as, for instance, in
\cite{bhattacharya2012nonparametric} and  \cite{tokdar2019bayesian}.
Indeed, as discussed in \cite{tokdar2019bayesian}, it is not even clear if
in such cases the need for an upper bound on the prior mass under the more
complex model is a natural requirement or rather a technical one. More
generally,  intuition about BFs (at least in parametric cases) is that
they tend to favor the more parsimonious model. In the particular context
described in Section \ref{sec:BF}, model $M_1$ can be regarded as a
degenerate case of model $M_2$, even though they are \virgolette{equally
complicated}. In this case, the above intuition evaporates, since
technically, embedding one model in the other is still one
infinite-dimensional model contained in another infinite-dimensional
model, and it is probably meaningless to ask which model is
\virgolette{simpler}. Under this scenario exploratory use of discrepancy
measures, such as those discussed in~\cite{gelman1996posterior}, may offer
some guidance.

In the simulation studies presented, our model always recovers the true
latent clustering among groups, thus providing empirical evidence in favor
 of our model to perform homogeneity tests.   We provide
some practical suggestions when the actual interest is on making this
decision. Our insight is that in order to prove asymptotic consistency of
the Bayes factor, one should introduce explicit separation between the
competing hypotheses. One possible way to accomplish this goal is, for
example, by introducing some kind of repulsion among the mixing measures
$F_i$'s in the model. This point will be focus of further study.

\section*{Acknowledgements}
     Fernando A. Quintana was supported by Fondecyt Grant 1180034. This work
     was supported by ANID - Millennium Science Initiative Program - NCN17\_059.

\appendix

\section{Proofs}\label{sec:s_proofs}

\underline{\textit{Proof of} {\bf Proposition \ref{prop:weak_supp}.}}\\
Consider $I=2$ for ease of exposition. We aim at showing that under
suitable choices of $G_0$ and $G_{00}$, the vector of random probability
measures $(G_1, G_2)$, where $G_i = F_{c_i}$ has full support on
$\Pp_\Theta \times \Pp_\Theta$.

This means that for every couple of distributions $(g_1, g_2) \in \Pp_\Theta
\times \Pp_\Theta$,  every weak neighborhood $W_1 \times W_2$ of $(g_1, g_2)$
receives non null probability.
In short, this condition entails $\pi_{\bm G}(W_1 \times W_2) > 0$.
Since $G_i = F_{c_i}$, we have that
\[
    \pi_{\bm G}(W_1 \times W_2) = \sum_{l, m=1}^2 \pi_{F_l, F_m}(W_1 \times W_2) \pi_c(l, m) > \pi_{F_1, F_2}(W_1 \times W_2) \pi_c(1, 2).
\]
Hence, since we are assuming that $\pi_c(l, m) > 0$ for all $l, m$, it is
sufficient to show that $\pi_{F_1, F_2}$, that is the measure associated
to the SemiHDP prior with $I=2$, has full weak support.

In the following, with a slight abuse of notation we denote by $\pi_{F_1,
F_2 \mid \Gtilde}(W_1 \times W_2)$ the measure associated to the SemiHDP
prior, conditional to a particular value of $\Gtilde$. We distinguish
three cases: $\kappa=1$, $0 < \kappa < 1$ and $\kappa=0$. The case
$\kappa=1$ is trivial, since $F_1$ and $F_2$ are marginally independently
distributed with Dirichlet process prior,  so that $\pi_{F_1, F_2}(W_1
\times W_2) = \mathcal{D}_{\alpha G_0}(W_1) \mathcal{D}_{\alpha G_0}(W_2)
> 0$ as long as $G_0$ has full support in $\Theta$ \citep[see, for
example,][]{ghosal2017fundamentals}.

Secondly consider $0 < \kappa < 1$, we show that as long as $G_0$ has full
support, then also $\pi_{\bm G}$ will have full support, regardless of the
properties of $G_{00}$. We have
\begin{multline}
 \pi_{F_1, F_2}(W_1 \times W_2) =  \int_{\Pp_\Theta} \pi_{F_1, F_2 \mid \Gtilde}(W_1 \times W_2) \Law(d\Gtilde) = \int_{\Pp_\Theta} \mathcal{D}_{\alpha \Ptilde}(W_1)
 \mathcal{D}_{\alpha \Ptilde}(W_2)  \Law(d\Gtilde).  \label{eq:supp_1}
\end{multline}
Now observe that if $G_0$ has full support, also $\Ptilde = \kappa G_0 +
(1-\kappa) \Gtilde$ will have full support, for any value of $\Gtilde$.
Hence by the properties of the Dirichlet Process, we get that $\pi_{F_1,
F_2}(W_1 \times W_2) > 0$ since the integrand in \eqref{eq:supp_1} is
bounded away from zero.

The case $\kappa=0$ is more delicate and requires additional work. We
follow the path outlined in \cite{deblasi2013asymptotic}, extending it to
our hierarchical case. In the following, let $\mathbb{S}^m$ denote the
$m-1$ dimensional simplex, i.e.
\[
    \mathbb{S}^m := \{(z_1, \ldots, z_m) \in \mathbb{R}^m \ : \ 0 \leq z_h \leq 1, \ h=1,\ldots ,m, \ \sum_{h=1}^m z_h = 1 \}
\]
Let $d_w$ denote the Prokhorov metric on $\Pp_\Theta$, which, as it is   well known,
metrizes the topology of the weak convergence on $\Pp_\Theta$.
Moreover, being $\Theta$ separable,  $(\Pp_\Theta, d_w)$ is separable as well
and the set of discrete measures with a finite number of point masses is
dense in $\Pp_\Theta$.

Hence, for any $(g_1, g_2)$ and any $\epsilon >0$, there exist two
discrete measures with weights $\bm p^{(i)} \in \mathbb{S}^{k_i}$ and
points $\bm x^{(i)} \in \Theta^{k_i}$ for $i=1, 2$ such that $d_w(F_{\bm
p^{(i)}, \bm x^{(i)}}, g_i) < \epsilon$, where $F_{\bm p^{(i)}, \bm
x^{(i)}} = \sum_k p^{(i)}_k \delta_{x^{(i)}_k}$. The difficulty when
$\kappa = 0$ is that conditionally on $\Gtilde$, the measure
$\mathcal{D}_{\alpha \Gtilde}$ does not have full weak support. Indeed,
its support is concentrated on the measures that have the same atoms of
$\Gtilde$. The proof will proceed as follows: start by defining weak
neighborhoods $W_i$ of $F_{\bm p^{(i)}, \bm x^{(i)}}$ by looking at
neighborhoods of their weights $\bm p^{(i)}$ ($U_i$) and atoms $\bm
x^{(i)}$ ($V_i$). Secondly, we join these neighborhoods.  If
$\Gtilde(\omega)$ belongs to this union (and this occurs with positive
probability), we guarantee that the atoms of both of $F_1$ and $F_2$, that
are shared with $\Gtilde$, are suited to approximate both $F_{\bm p^{(i)},
\bm x^{(i)}}$, $i=1,2$. Hence, by the properties of the Dirichlet Process
one gets the support property.

More in detail, define the sets
\begin{align*}
    V_i(\delta) = \{\bm x_i \in \Theta^{k_i} \ s.t. \ |x_{ij} - x^{(i)}_j| < \delta  \}, i=1,2
\end{align*}
and let $V = V_1 \cup V_2$. Then we operate a change of index by
concatenating $\bm x^{(1)}$ and $\bm x^{(2)}$, and call it $\bm x^*$, i.e.
$\bm x^* = [\bm x^{(1)}, \bm x^{(2)}]$. Hence we characterize the set $V$
as
\[
    V(\delta) = \{\bm x \in \Theta^{k_1 + k_2} \ s.t \ |x_j - x^*_j| < \delta \}.
\]
Secondly, define $\bm p^*$ by concatenating $\bm p^{(1)}$ and
and $\bm p^{(2)}$:  $\bm p^* = [\bm p^{(1)}, \bm p^{(2)}]$ and let
\begin{align*}
    U_1(\eta) &= \{ \bm p \in \mathbb{S}^{k_1 + k_2} \ s.t. \ |p_{j} - p^*_j| < \eta \text{ for } j=1, \ldots, k_1, \ |p_{j} - 0| < \eta \text{ elsewhere}  \} \\
    U_2(\eta) &= \{ \bm p \in \mathbb{S}^{k_1 + k_2} \ s.t. \ |p_{j} - p^*_j| < \eta \text{ for } j=k_1 + 1, \ldots, k_1 + k_2, \ |p_{j} - 0| < \eta \text{ elsewhere}  \}
\end{align*}
Finally, define the following neighborhoods
\begin{align*}
    W_i &:= \{\sum_{j=1}^{k_1 + k_2} p_{j} \delta_{x_j} \text{ for any } \bm p \in U_i, \text{ and any }  \bm x \in V \}, i=1,2 \\
    W_0 &:= \{\sum_{j=1}^{k_1 + k_2} p_{j} \delta_{x_j} \text{ for any } \bm p \in \mathbb{S}^{k_1 + k_2}, \text{ and any }  \bm x \in V \}.
\end{align*}
This means that the $V_i$ sets  are the neighborhoods of the atoms $\bm
x_i$ that are well suited to approximate $F_{\bm p^{(i)}, \bm x^{(i)}}$
and $V$ is their union. The sets $U_i$, $i=1,2$, instead, are related to
the weights of  $F_{\bm p^{(i)}, \bm x^{(i)}}$. In particular, each $U_i$
is constructed in such a way to approximate well $\bm p^{(i)}$ (a vector
in $\mathbb{S}^{k_i}$) with a vector of weights in $\mathbb{S}^{k_1 +
k_2}$. This is necessary because if $\Gtilde$ has support points in $V$,
so will do the draws $F_1$ and $F_2$ from $\mathcal{D}_{\alpha \Gtilde}$.
However, by assigning  a negligible weight in $U_1$ to the atoms $\bm
x^{(2)}$ and vice-versa for the atoms $\bm x^{(1)}$ in $U_2$, we guarantee
that the probability measures in $W_i$ constitute a weak neighborhood of
$F_{\bm p^{(i)}, \bm x^{(i)}}$ for each $i=1,2$.

From \cite{deblasi2013asymptotic}, it is sufficient to show that
$\pi_{F_1, F_2}(W_1 \times W_2) > 0$ since for appropriate choices of
$\eta$ and $\delta$ one has that $d_w(\widetilde F_1, g_1) + d_w(
\widetilde F_2, g_2) < \epsilon$ for all choices of $\widetilde  F_1 \in
W_1$ and $\widetilde  F_2 \in W_2$. Hence
\begin{align*}
    \pi_{F_1, F_2}(W_1 \times W_2) = \int_{\Pp_\Theta} \pi_{F_1, F_2 \mid \Gtilde}(W_1 \times W_2) \Law(d \Gtilde)  &\geq \int_{W_0} \pi_{F_1, F_2 \mid \Gtilde }(W_1, W_2) \Law(d\Gtilde) \\
&= \int_{W_0} \mathcal{D}_{\alpha \Gtilde}(W_1) \mathcal{D}_{\alpha \Gtilde}(W_2) \Law(d\Gtilde)
\end{align*}
Now observe that for any $\Gtilde \in W_0$, we have that
$\mathcal{D}_{\alpha \Gtilde}(W_i) > 0$. This follows again from the
properties of the Dirichlet process, since for any value of
$\Gtilde(\omega)$, there exists a non-empty set $\widetilde W_i \subset
W_i$, $\widetilde W_i = \{\widetilde F_i \in W_i: \ supp(\widetilde F_i)
\subset supp(\Gtilde)\}$. Hence $\pi_{F_1, F_2 \mid \Gtilde}(W_1 \times
W_2) \geq \pi_{F_1, F_2 \mid \Gtilde}(\widetilde W_1 \times \widetilde
W_2) > 0$, since the Dirichlet process gives positive probability to the
weak neighborhoods of measures whose support is contained in the support
of its base measure, i.e. $\Gtilde$. \qed

\bigskip
\noindent
\underline{\textit{Proof of} {\bf Covariance of the semi-HDP}}.\\
If  $(F_1,F_2) \sim semiHDP(\alpha, \gamma, \kappa, G_0,
G_0)$, then
\begin{align*}
\cov(F_1(A) F_2(B)) &= \E \left[F_1(A) F_2(B) \right] - \E \left[F_1(A) \right] \E \left[F_2(B) \right] \\
&= \E \left[ \E \left[F_1(A) F_2(B) \mid \Ptilde \right] \right] - \E \left[ \E \left[F_1(A) \mid \Ptilde \right] \right] \E \left[ \E \left[F_2(B) \mid \Ptilde \right] \right] \\
&= \E \left[ \E \left[F_1(A) \mid \Ptilde \right]\E \left[F_2(B) \mid \Ptilde \right] \right] - G_0(A) G_0(B)\\
&= \E \left[\Ptilde(A) \Ptilde(B) \right] - G_0(A) G_0(B) \\
&= \kappa^2 G_0(A) G_0(B) + \kappa(1-\kappa) G_0(A) \E\left[\Gtilde(B)\right] + \kappa(1-\kappa) G_0(B) \E\left[\Gtilde(A)\right] \\
& \quad +  (1 - \kappa)^2 \E \left[\Gtilde(A) \Gtilde(B) \right] - G_0(A)G_0(B)\\
&= (1 - \kappa)^2 \E \left[\Gtilde(A) \Gtilde(B) \right] - (1-\kappa)^2 G_0(A)G_0(B)   \\
&= (1-\kappa)^2 \cov(\Gtilde(A), \Gtilde(B))
= \frac{(1 - \kappa)^2}{1 + \gamma} \left(G_0(A \cap B) - G_0(A)G_0(B) \right).
\end{align*}
The last equality follows because $\Gtilde$ is a Dirichlet process.
\qed

\clearpage
\noindent
\underline{\textbf{Higher order moments}}.\\
To compute higher order moments,   we make use of a result from~\cite{argiento2019hcrm}.
Let $F_1 \mid \Ptilde \sim \mathcal{D}_{\alpha\Ptilde}$ as in
\eqref{eq:iid_DP} - \eqref{eq:hdp_g0}; then one has, for any set $A \in \mathcal{B}(\Theta)$:
\[
   \E[F_1(A)^n \mid \Gtilde] = \sum_{t=1}^n \Ptilde(A)^t P(K_n = t),
\]
where $K_n$ is the random variable representing the number of
\emph{clusters} in a sample of size $n$;  see (15)  in
\cite{argiento2019hcrm}. If, as in our case, the base measure is not
absolutely continuous, the term clusters might be misleading as they do
not coincide with the unique values in the sample, but rather with the
number of the tables in the Chinese restaurant process. In the following
we refer to cluster or table interchangeably. Hence, we have:
\begin{align*}
	\E[F_1(A)^n] &= \E[\E[F_1(A)^n \mid \Gtilde]] = \E \left[\sum_{t=1}^n \Ptilde(A)^t P(K_n = t) \right]\\
	&= \E \left[\sum_{t=1}^n P(K_n = t) \sum_{h=0}^t \binom{t}{h} (\kappa G_0(A))^{t-h} \times ((1-\kappa) \Gtilde(A))^h \right] \\
	&= \sum_{t=1}^n P(K_n = t) \sum_{h=0}^t \binom{t}{h} (\kappa G_0(A))^{t-h} (1-\kappa)^h \E[\Gtilde(A)^h] \\
	&= \sum_{t=1}^n P(K_n = t) \sum_{h=0}^t \binom{t}{h} (\kappa G_0(A))^{t-h} (1-\kappa)^h \sum_{m=1}^h G_{00}(A) P(\widetilde K_h = m),
\end{align*}
where $\widetilde K_h$ is the number of clusters from a sample of size $h$ from the DP $\Gtilde$. Moreover, if we assume $G_0 = G_{00}$  we get
\begin{equation*}
\E[F_1(A)^n]  = \sum_{t=1}^n P(K_n = t) \sum_{h=0}^t \binom{t}{h} \kappa^{t-h} (1-\kappa)^h \sum_{m=1}^h G_0(A)^{t-h + m} P(K_h = m).
\end{equation*}

\begin{figure}
	\centering
	\includegraphics[scale=0.7]{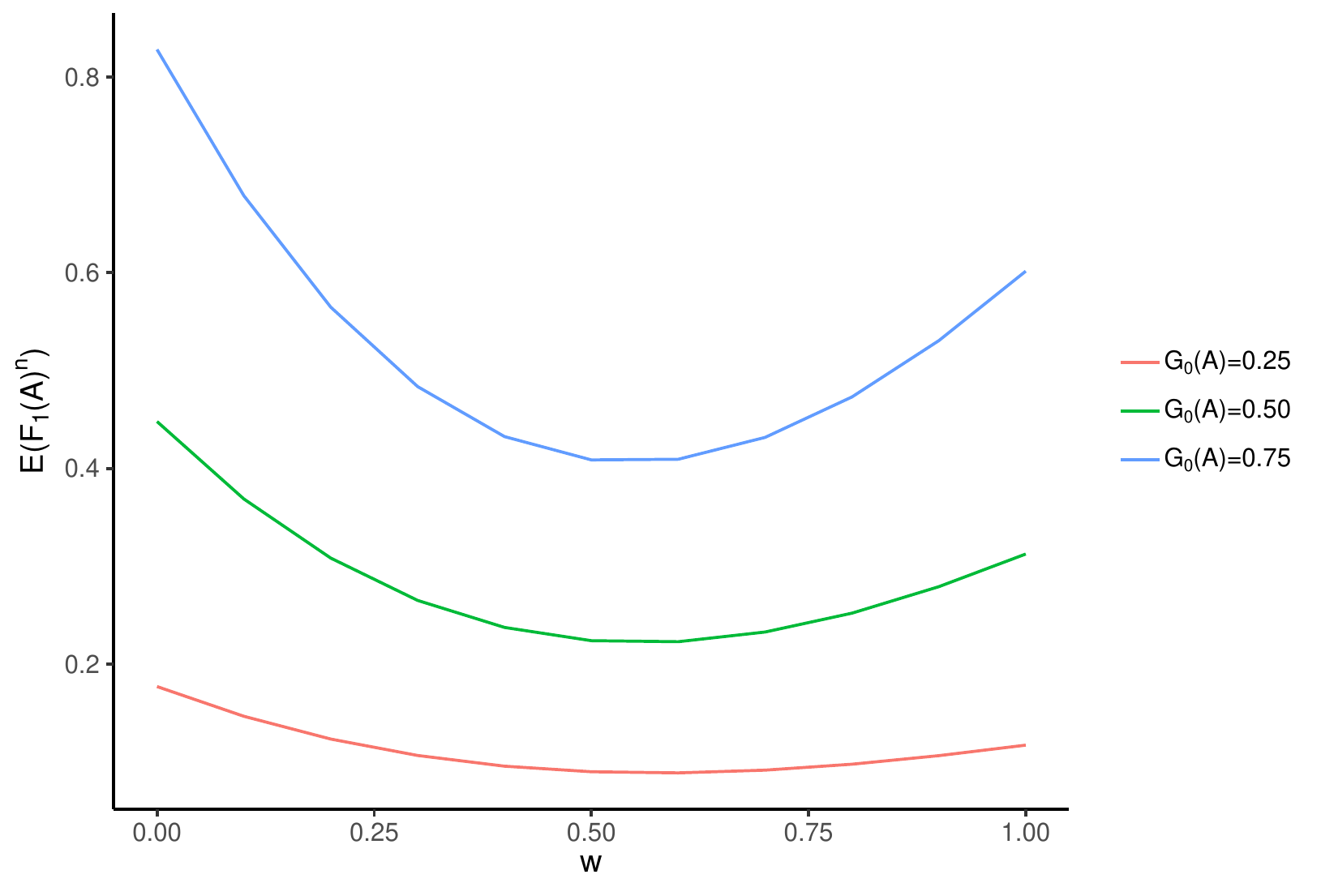}
	\caption{3-rd moment of $F_1(A)$ for increasing values of $\kappa$ and various values of $G_0(A)$.}
	\label{fig:nth_moment}
\end{figure}

Figure~\ref{fig:nth_moment} shows the effect of the parameter $\kappa$
over $\E[F_1(A)^3]$ for various values of $G_0(A)$. The limiting cases of
the standard Dirichlet process and the Hierarchical Dirichlet Process are
recovered when $\kappa=1$ and $\kappa=0$ respectively.

\bigskip
\noindent \underline{\textit{Proof of} {\bf Proposition
\ref{prop:degeneracy}}}.\\
Indicating with $\tau_j$ the shared unique values between $\bm \theta_1$
and $\bm \theta_2$, and with $\theta^*_{ij}$ the unique values in sample
$\bm \theta_i$ that are specific to group $i$, i.e. not shared, the pEPPF,
given $\bm c$, can be written as:
\begin{equation*}
	\Pi^{(N)}_k (\bm n_1, \bm n_2, \bm q_1, \bm q_2 | \bm c) = \int_{\Theta^k} \mathbb{E} \left[\prod_{j=1}^{k_1}
F_{c_1}^{n_{1j}} (d\theta^*_{1j})  \prod_{j=1}^{k_2} F_{c_2}^{n_{2j}} (d\theta^*_{2j})  \prod_{j=1}^{k_0} F_{c_1}^{q_{1j}} (d\tau_{j}) F_{c_2}^{q_{2j}} (d\tau_{j}) \right].
\end{equation*}
See (23) in \cite{Cam_etal_2018latent}. Marginalizing out $\bm c$ we obtain that:
\begin{align*}
\Pi^{(N)}_k (\bm n_1, \bm n_2, \bm q_1, \bm q_2) = \sum_{l,m = 1}^2 \pi_{\bm c}(\bm c = (l,m)) \Pi^{(N)}_k (\bm n_1, \bm n_2, \bm q_1, \bm q_2 | \bm c = (l, m)).
\end{align*}
The cases $\bm c = (1, 1)$ and $\bm c = (2, 2)$ can be easily managed as
it corresponds to full exchangeability and the EPPF corresponding to those
cases is already available. Hence, let us consider the case when $\bm c =
(1, 2)$, as the case $\bm c=(2,1)$ will be identical because the $F_i$'s
are iid.
\begin{align*}
\Pi^{(N)}_k &(\bm n_1, \bm n_2, \bm q_1, \bm q_2 | \bm c = (1, 2)) = \int_{\Theta^k} \mathbb{E}
 \left[\prod_{j=1}^{k_1} F_{1}^{n_{1j}} (d\theta^*_{1j})  \prod_{j=1}^{k_2} F_{2}^{n_{2j}} (d\theta^*_{2j})  \prod_{j=1}^{k_0} F_{1}^{q_{1j}} (d\tau_{j}) F_{2}^{q_{2j}} (d\tau_{j}) \right]  \\
&= \int_{\Theta^k}
	\mathbb{E} \left[\prod_{j=1}^{k_1} F_{1}^{n_{1j}} (d\theta^*_{1j}) \prod_{j=1}^{k_0} F_{1}^{q_{1j}}(d\tau_{j}) \right]
	\mathbb{E} \left[ \prod_{j=1}^{k_2} F_{2}^{n_{2j}} (d\theta^*_{2j})  \prod_{j=1}^{k_0} (d\tau_{j}) F_{2}^{q_{2j}} (d\tau_{j}) \right]
\end{align*}
since $F_1$ and $F_2$ are independent. The first expected value is the
joint probability of $\Pi_{k_1 + k_0}^{N_1}$ (the EPPF of a partition of
$N_1$ objects into $k_1 + k_0$ groups with vectors of frequencies $\bm
n_1, \bm q_1$) and the set of unique values is denoted by $(x_{11}, \dots,
x_{1k_1}, \tau_1, \dots \tau_{k_0})$. Similarly for the second expected
value. Because  $F_1 \sim \mathcal{D}_{\alpha G_0}$,  we can rewrite the
expected value as:
\begin{align*}
&\mathbb{E} \left[\prod_{j=1}^{k_1} F_{1}^{n_{1j}} (d\theta^*_{1j}) \prod_{j=1}^{k_0} F_{1}^{q_{1j}}(d\tau_{j}) \right]\\
& \quad = \frac{\alpha_1^{k_1 + k_0} \Gamma(\alpha_1)}{\Gamma(\alpha_1 + N_1)} \prod_{j=1}^{k_1} \Gamma(n_{1j}) \prod_{j=1}^{k_0} \Gamma(q_{1j})
\prod_{j=1}^{k_1} G_0 (d\theta^*_{1j}) \prod_{j=1}^{k_0} G_0 (d\tau_j).
\end{align*}
Hence, we have that
\begin{align*}
\Pi^{(N)}_k &(\bm n_1, \bm n_2, \bm q_1, \bm q_2 | \bm c = (1, 2)) = \\
&\quad =
\int_{\Theta^k}
	\mathbb{E}  \left[\prod_{j=1}^{k_1} F_{1}^{n_{1j}} (d\theta^*_{1j}) \prod_{j=1}^{k_0} F_{1}^{q_{1j}}(d\tau_{j}) \right]
	\mathbb{E} \left[ \prod_{j=1}^{k_2} F_{2}^{n_{2j}} (d\theta^*_{2j})  \prod_{j=1}^{k_0} (d\tau_{j}) F_{2}^{q_{2j}} (d\tau_{j}) \right] \label{eq:cond_peppf} \\
&\quad = \frac{\alpha_1^{k_1 + k_0} \Gamma(\alpha_1)}{\Gamma(\alpha_1 + N_1)}
\frac{\alpha_2^{k_2 + k_0} \Gamma(\alpha_2)}{\Gamma(\alpha_2 + N_2)}
\prod_{j=1}^{k_1} \Gamma(n_{1j}) \prod_{j=1}^{k_2} \Gamma(n_{2j}) \prod_{j=1}^{k_0} \Gamma(q_{1j} ) \Gamma(q_{2j}) \notag \\
&\qquad\qquad\qquad\times\int_{\Theta^k} \prod_{j=1}^{k_1} G_0 (d\theta^*_{1j}) \prod_{j=1}^{k_2} G_0 (d\theta^*_{2j}) \prod_{j=1}^{k_0} G_0 (d\tau_j) G_0 (d\tau_j). \notag
\end{align*}

Looking at the last integral, we can see that this is clearly 0 unless
$k_0 = 0$, in fact, consider $k_0 = 1$:
\begin{align*}
\int_{\Theta^{k-1}} \prod_{j=1}^{k_1} G_0 (d\theta^*_{1j}) \prod_{j=1}^{k_2} G_0 (d\theta^*_{2j}) \int_{\Theta} G_0 (dz) G_0 (dz)
\end{align*}
and observe that the last integral is integrating the product measure
$G_0\times G_0$ on the straight line $y=x$, resulting thus in 0.

Summing up, if $k_0 = 0$ we get:
\begin{align*}
	\Pi^{(N)}_k  (\bm n_1, \bm n_2, \bm q_1, \bm q_2)=
	& \pi_1 \frac{\alpha^{k_1 + k_2} \Gamma(\alpha)}{\Gamma(\alpha + N)} \prod_{j=1}^{k_1} \Gamma(n_{1j}) \prod_{j=1}^{k_2} \Gamma(n_{2j})   \\
	 & + (1 - \pi_1) \frac{\alpha^{k_1 + k_2} \Gamma(\alpha)^2}{\Gamma(\alpha + N_1) \Gamma(\alpha + N_2)}
\prod_{j=1}^{k_1} \Gamma(n_{1j}) \prod_{j=1}^{k_2} \Gamma(n_{2j})
\end{align*}
else, if $k_0 > 0$:
\begin{align*}
	\Pi^{(N)}_k & (\bm n_1, \bm n_2, \bm q_1, \bm q_2) = \\
	& (1-\pi_1) \frac{\alpha^{k_1 + k_2 + k_0} \Gamma(\alpha)}{\Gamma(\alpha + N)} \prod_{j=1}^{k_1} \Gamma(n_{1j}) \prod_{j=1}^{k_2} \Gamma(n_{2j}) \prod_{j=1}^{k_0} \Gamma(q_{1j} + q_{2j})
\end{align*}
which can be rewritten down as in \cite{Cam_etal_2018latent}; call
\[
 \Phi_k^{(N)}(\bm n_1, \bm n_2, \bm q_1 + \bm q_2) = \frac{\alpha^{k_1 + k_2 + k_0} \Gamma(\alpha)}{\Gamma(\alpha + N)} \prod_{j=1}^{k_1} \Gamma(n_{1j}) \prod_{j=1}^{k_2} \Gamma(n_{2j}) \prod_{j=1}^{k_0} \Gamma(q_{1j} + q_{2j})
\]
the EPPF of the fully exchangeable case, and
\[
 \Phi_{k_0 + k_i}^{(N_i)} (\bm n_i, \bm q_i) = \frac{\alpha^{k_i + k_0} \Gamma(\alpha)}{\Gamma(\alpha + N_i)} \prod_{j=1}^{k_i} \Gamma(n_{ij}) \prod_{j=1}^{k_0} \Gamma(q_{ij})
\]
the marginal EPPF for the individual groups $i=1, 2$.
We have that:
\begin{equation*}
	\Pi^{(N)}_k  (\bm n_1, \bm n_2, \bm q_1, \bm q_2) = \pi_1 \Phi_k^{(N)}(\bm n_1, \bm n_2, \bm q_1 + \bm q_2) + (1 - \pi_1) \Phi_{k_0 + k_1}^{(N_1)} (\bm n_1, \bm q_1) \Phi_{k_0 + k_1}^{(N_2)} (\bm n_2, \bm q_2) I(k_0 = 0)
\end{equation*}
which is \eqref{eq:peppf_deg}.
\qed

\bigskip
\noindent \underline{\textit{Proof of} {\bf Proposition
\ref{prop:marginal_law}}}.\\
Of course, the marginal law of $(\bm \theta_1, \ldots, \bm \theta_I)$,
conditional to $\bm c$, can be computed as
\begin{align*}
	\Law &(d \bm \theta_1, \ldots d \bm \theta_I \mid \bm c) = \int_{\Pp_\Theta} \ldots \int_{\Pp_\Theta} \Law (d \bm \theta_1, \ldots d \bm \theta_I \mid F_1, \ldots F_I, \bm c) \Law(dF_1, \ldots dF_I).
\end{align*}
Now we operate a change of indices and call $\bm \theta_{r} = \{\bm
\theta_i = (\theta_{i1}, \ldots \theta_{iN_i}): c_i  = r\}$, so that $(\bm
\theta_1, \ldots \bm \theta_I) = (\bm \theta_{r_1}, \ldots \bm
\theta_{r_R})$ where $R$ is the number of unique values in $\bm{c}$, i.e.
the number of non-empty restaurants. We get
\begin{align*}
	\Law (d \bm \theta_1, \ldots d \bm \theta_I \mid \bm c)
		  &= \int_{\Pp_\Theta} \int_{\Pp_\Theta} \ldots \int_{\Pp_\Theta} \Law (d \bm \theta_{r_1}, \ldots d \bm \theta_{r_R} \mid F_1, \ldots F_I, \bm c) \Law(dF_1, \ldots dF_I \mid \Gtilde) \Law(d\Gtilde) \\
		 &= \int_{\Pp_\Theta} \left( \prod_{i=1}^R \int_{\Pp_\Theta} \Law(d \bm \theta_{r_i} \mid F_{r_i}) \Law (d F_{r_i} \mid \Gtilde) \right) \Law(d \Gtilde).
\end{align*}
Observe that
\[
	\int_{\Pp_\Theta} \Law(d \bm \theta_{r_i} \mid F_{r_i}) \Law(dF_{r_i} \mid \Gtilde)= \Law (\rho_{r_i}) \prod_{j=1}^{H_{r_i}} \Ptilde(d \theta^*_{r_i j}),
\]
where $\rho_{r_i}$ is the partition induced by the $\ell$-clusters in the
$r_i$ restaurant. We use the same definition of $\ell$-cluster as in
\cite{argiento2019hcrm}. We underline that $\{ \theta^*_{r_i j}, j=1,
\ldots, H_{r_i} \}$ are not the unique values in the sample, since the
base measure is atomic. Hence we have
\begin{align*}
&\int_{\Pp_\Theta} \left( \prod_{i=1}^R \int_{\Pp_\Theta} \Law(d \bm \theta_{r_i} \mid F_{r_i}) \Law (d F_{r_i} \mid \Gtilde) \right) \Law(d \Gtilde) \\&%
= \left(\prod_{i=1}^{R} \Law (\rho_{r_i}) \right)\int_{\Pp_\Theta} \prod_{i=1}^{R} \prod_{j=1}^{H_{r_i}} \Ptilde (d \theta^*_{r_i j}) \Law(d \Gtilde).
\end{align*}
Now observe how the values $\{\theta^*_{r j}: \ r=1, \ldots R, j=1, \ldots
H_{r_i}\}$ are all iid from $\Ptilde$. So, there is no need for the
division into restaurants anymore. We can thus stack all the vectors $\bm
\theta^*_{r_i}$ together, apply a change of indices $(r_i, j) \rightarrow
l$ so that now these $\{\theta^*_{r_i}\}$ are represented by $(\theta^*_1,
\ldots, \theta^*_L)$ and
\begin{align*}
	\Law (d \bm \theta_1, \ldots d \bm \theta_I \mid \bm c) &=\prod_{i=1}^{R} \Law (\rho_{r_i}) \int_{\Pp_\Theta} \prod_{l=1}^{L} \Ptilde (d \theta^*_{l}) \Law(d \Gtilde) \\
	 &= \prod_{i=1}^{R} \Law (\rho_{r_i}) \int_{\Pp_\Theta} \prod_{l=1}^{L} \left( \kappa G_0 (d \theta^*_{l}) + (1-\kappa) \Gtilde(d \theta^*_{l}) \right) \Law(d \Gtilde).
\end{align*}	
Now, as done in Section \ref{sec:model}, we introduce a set of latent variables $\bm h = (h_1, \ldots, h_L)$, $h_l \iid \text{Bernoulli}(\kappa)$, that gives
\begin{align*}
	\Law &(d \bm \theta_1, \ldots d \bm \theta_I \mid \bm c) = \prod_{i=1}^{R} \Law(\rho_{r_i}) \sum_{\bm h \in \{0, 1\}^L} p(\bm h) \int_P \prod_{l=1}^L G_0(d\theta^*_l)^{h_l} \times \Gtilde(d\theta^*_l)^{1-h_l} \Law(d \Gtilde) \\
	&= \prod_{i=1}^{R} \Law(\rho_{r_i}) \sum_{\bm h \in \{0, 1\}^L} p(\bm h) \prod_{l=1}^L G_0(d\theta^*_l)^{h_l} \int_P \prod_{l=1}^L \Gtilde(d\theta^*_l)^{1-h_l} \Law(d \Gtilde) \\
	&= \prod_{i=1}^{R} \Law(\rho_{r_i}) \sum_{\bm h \in \{0, 1\}^L} p(\bm h) \prod_{l=1}^L G_0(d\theta^*_l)^{h_l} \times \Law(\eta \mid \bm h) \prod_{k=1}^{M(\eta)} G_{00}(d \theta^{**}_k),
\end{align*}
where $\eta$ is the partition of the $\{\theta^*_l : \ l=1, \ldots, L
\text{ and }h_l = 0\}$, i.e. the partition of $\sum_{l=1}^L (1-h_l)$
objects arising form the Dirichlet process $\Gtilde$, while
$\{\theta^{**}_k\}$ are the unique values among $\{\theta^*_l : \ l=1,
\ldots, L \text{ and }h_l = 0\}$ and $p(\bm h) = \prod_{l=1}^L
\kappa^{h_l} (1-\kappa)^{1-h_l}$ is the joint distribution of $\bm h$.
\qed

\bigskip
\noindent \underline{\textit{Proof of} {\bf Proposition
\ref{prop:consistency}}}.\\
Model $M_2$ defines a prior $\Pi_2$ on the space of densities $(p, q) \in
\Pp_\Y \times \Pp_\Y$. On the other hand, model $M_1$ defines a prior on
$\Pp_\Y$. However, by embedding $\Pp_\Y$ in the product space $\Pp_\Y
\times \Pp_\Y$ via the mapping $p \mapsto (p, p)$, we can  also consider
the prior $\Pi_1$ induced by model $M_1$ as a measure on (a subset of)
$\Pp_\Y \times \Pp_\Y$.

Now, showing that $\Pi_2$ satisfies the Kullback-Leibler property is a
straightforward application of Theorem 3 in \cite{wu2008kullback}, under
the same set of assumptions on the kernel $k(\cdot|\theta)$, and on $p_0$
and $q_0$, that we do not report here. Notice that these assumptions are
satisfied when $k(\cdot|\theta)$ is the univariate Gaussian kernel with
parameters given by the mean and the scale, and under standard  regularity
conditions on $p_0$ and $q_0$.

Now we turn our attention to $\Pi_1$. It is obvious to argue that $\Pi_1$
does not have the Kullback-Leibler property in the larger space $\Pp_\Y
\times \Pp_\Y$, since it gives positive mass only to sets $\{(p, q) \in
\Pp_\Y \times \Pp_\Y: p=q \}$. Consequently,  if $p_0 \neq q_0$, one will
have that for a small enough $\delta$:
\[
    \Pi_1\left((p, q): D_{KL}((p, q), (p_0, q_0) < \delta \right) = 0,
\]
thus proving that $\Pi_1$ does not have the Kullback-Leibler
property.

In summary, under the same assumptions on $p_0, q_0$  and the kernel
$k(\cdot \mid \theta)$ as in \cite{ghosal2008nonparametric}, and assuming
$p_0 \neq q_0$, we are comparing a model ($M_2$) with the Kullback-Leibler
property against one ($M_1$) that does not have it. Theorem 1 in
\cite{walker2004priors} implies that the Bayes factor consistency is
ensured. \qed

\bigskip
\noindent \underline{\textit{Proof of} {\bf Equation
\eqref{eq:mix_dist}}}.\\
Let $p_r = \sum_{i=1}^{H_r} w_{ri} \calN(\mu_{ri}, \sigma^2_{ri})$ and $p_m =
\sum_{j=1}^{H_m} w_{mj} \calN(\mu_{mj}, \sigma^2_{mj})$ be the mixture densities associated to the mixing measures $F_r$ and $F_m$ respectively. Observe that both $H_m$ and $H_r$ are finite here as $F_r$ and $F_m$ have been approximated as shown in the description of the Gibbs sampler in Section~\ref{sec:MCMC}.
Then
\begin{align*}
	d^2(F_r, F_m) &= L_2^2(p_r, p_m) = \int (p_r(y) - p_m(y))^2 dy \\
	& = \int \left( \sum_{i=1}^{H_r} w_{ri} \calN(y; \mu_{ri}, \sigma^2_{ri})  - \sum_{j=1}^{H_m} w_{mj} \calN(y; \mu_{mj}, \sigma^2_{mj})\right)^2 dy
\end{align*}
For any value of $y$ the above integrand reduces to 
\begin{multline*}
  \Bigg( \sum_{i=1}^{H_r} w_{ri} \calN(y; \mu_{ri}, \sigma^2_{ri}) \Bigg) ^2 + \Bigg( \sum_{j=1}^{H_m} w_{mj} \calN(y; \mu_{mj}, \sigma^2_{mj}) \Bigg) ^2 + \\
    \qquad\qquad \qquad -2 \Bigg(\sum_{i=1}^{H_r} w_{ri} \calN(y; \mu_{ri}, \sigma^2_{ri}) \Bigg)\Bigg(\sum_{j=1}^{H_m} w_{mj} \calN(y; \mu_{mj}, \sigma^2_{mj})\Bigg)
\end{multline*}
Each term in the right hand side can be expressed as a product of two summations, say $(\sum_{i} a_i)(\sum_j b_j) = \sum_{i, j} a_i b_j$. When $\{a_i\}$ and $\{b_j\}$ are equal, this further reduces to $\sum_{i, i^\prime} a_i a_{i^\prime}$.

Hence, exchanging summations and integrals, $d^2(F_r, F_m)$ equals
\begin{align*}	
	 d^2(F_r, F_m) & =\sum_{i, i^\prime =1}^{H_r} w_{ri}, w_{ri^\prime} \int \calN(y;\mu_{ri}, \sigma^2_{ri}) \calN(y;\mu_{ri^\prime}, \sigma^2_{ri^\prime}) dy  \\
	& \qquad\qquad \qquad
	+\sum_{j, j^\prime=1}^{H_m} w_{mj}, w_{mj^\prime} \int \calN(y;\mu_{mj}, \sigma^2_{mj}) \calN(y;\mu_{mj^\prime}, \sigma^2_{mj^\prime}) dy
	\\
	& \qquad\qquad \qquad -2 \sum_{i=1}^{H_r} \sum_{j=1}^{H_m} w_{ri} w_{mj} \int \calN(y; \mu_{ri}, \sigma^2_{ri}) \calN(y;\mu_{mj}, \sigma^2_{mj}) dy.
\end{align*}

\bigskip
\noindent \underline{\textit{Proof of} {\bf Equation
\eqref{eq:normal_prod_int}}}.\\
This follows immediately from Equation (371) in \cite{matrixcookbook}.

\section{Discussion of Bayes Factor consistency in the homogeneous case}
\label{sec:s_bf} When $p_0 = q_0$, consistency of the Bayes factor would
require $BF_{12} \rightarrow +\infty$.  This is a result we have not been
able to prove so far, but it is worth pointing out the following relevant
issues. To begin with, note that both models $M_1$ and $M_2$ have   the
Kullback-Leibler property. Several papers discuss this case, for example
Corollary 3.1 in \cite{ghosal2008nonparametric}, Section 5 in
\cite{chib2016bayes} and Corollary 3 in \cite{chatterjee2020short} in the
general setting of dependent data. For more specific applications, refer
also to \cite{tokdar2019bayesian} where the focus is on testing
Gaussianity of the data under a Dirichlet process mixture alternative,
\cite{mcvinish2009bayesian}  for goodness of fit tests using mixtures of
triangular distribution  and \cite{bhattacharya2012nonparametric} for data
distributed over non-euclidean manifolds.

As pointed out in \cite{tokdar2019bayesian}, the hypotheses in Corollary
3.1 by \cite{ghosal2008nonparametric} are usually difficult to prove,
since they require a lower bound on the prior mass $\Pi_2$ around
neighborhoods of $(p_0, p_0)\in \mathbb{P}_{\mathbb{Y}}\times
\mathbb{P}_{\mathbb{Y}}$. To the best of our knowledge, this kind of
bounds have been derived only for the very special kind of mixtures in
\cite{mcvinish2009bayesian}. Similarly,  the approach by
\cite{chib2016bayes} would require a knowledge of such lower bounds too
(see for instance their Assumption 3). Corollary 3 in
\cite{chatterjee2020short} does not apply in our case as well, because one
of their main assumptions presumes that both models specify a population
distribution (i.e. a likelihood) with density w.r.t some {\em common}
$\sigma$--finite measure, together with the true distribution of the data.
In our case $M_1$ specifies random probability measures that are
absolutely continuous w.r.t the Lebesgue measure on $\mathbb{R}$, while
under model $M_2$ the random probability measures have density under the
Lebesgue measure on $\mathbb{R}^2$.

\section{Relabeling step}\label{sec:s_relabeling}

\begin{figure}[ht]
    \centering
    \includegraphics[scale=1.5]{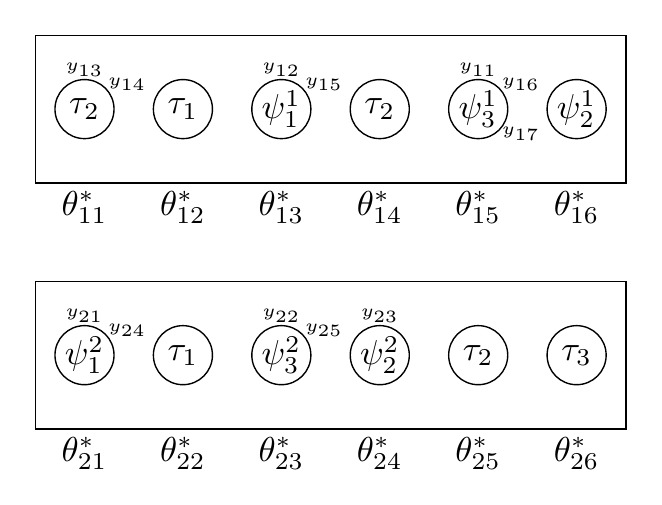}
    \caption{The state at one particular iteration}
    \label{fig:relabel}
\end{figure}

In the following, we adopt a slightly different notation to simplify the
pseudocode notation. Figure~\ref{fig:relabel} depicts the state at a
particular iteration. We denote by $\psi_{rh}$ the atoms in restaurant $r$
arising from $G_0$ and with $\tau_h$ the atoms arising from $G_{00}$.
Observe how in restaurant 1 the value $\tau_2$ appears more than once.

In our implementation, the state composed by $\bm \psi_{r}$, $\bm \tau$
(i.e. all the unique values of the atoms) and the indicator variables
$\{t_{rl}\}$ and $\{h_{rl}\}$ that let us reconstruct the value of
$\theta^*_{rl}$. In particular if $\theta_{rl} = \psi_{rk}$ if $h_{rl}=1$
and $t_{rl} = k$. Instead $\theta_{rl} = \tau_m$ if $h_{rl} = 0$ and
$t_{rl} = m$. Moreover we also have the latent variables $s_{ij}$ as
 described in  Equation \eqref{eq:cluster_alloc}.

For the example in Figure~\ref{fig:relabel}, the  latent variables assume
the following values for the first restaurant
\begin{align*}
    \bm s_1 = [5, 3, 1, 1, 3, 5, 5] \qquad
    \bm h_1 = [0, 0, 1, 1, 0, 0] \qquad
    \bm t_1 = [2, 1, 1, 3, 3, 3]
\end{align*}
while for the second restaurant
\begin{align*}
    \bm s_2 = [1, 3, 4, 1, 3] \qquad
    \bm h_2  = [1, 0, 1, 1, 0, 0] \qquad
    \bm t_2  = [1, 1, 3, 2, 2, 3]
\end{align*}

During the relabeling step, we look at the number of customers in each
table and find out that $\theta^*_{12}, \theta^*_{14}, \theta^*_{16},
\theta^*_{22}, \theta^*_{25}$ and $\theta^*_{26}$ are not used. Moreover
also $\tau_1, \tau_3$ and $\psi^1_2$ are not used.

This leads to the following relabel
\begin{align*}
    \bm s_1^{new}  = [3, 2, 1, 1, 2, 3, 3] \qquad
    \bm h_1^{new}  = [0, 1, 1] \qquad
    \bm t_1^{new}  = [1, 1, 2]
\end{align*}
and
\begin{align*}
    \bm s_2^{new}  = [1, 2, 3, 1, 2] \qquad
    \bm h_2^{new}  = [1, 1, 1] \qquad
    \bm t_2^{new}  = [1, 3, 2]
\end{align*}
In the code, the transformation $\bm s_i \rightarrow \bm s_i^{new}$ is
straightforward. Moreover $\bm h_i^{new}$ is computed from $\bm h_i$ by
selecting only the elements corresponding to the sorted unique values in
$\bm s_i$. For example the unique values in $\bm s_i$ are $[1, 3, 5]$ and
$\bm h_i^{new} = \left[\bm h_i[1], \bm h_i[3], \bm h_i[5] \right]$.

The only complicated step is the one concerning $\bm t$. To update this
last set of indicator variables we build two maps: $\tau_{map}$ and
$\psi_{map}$ that associate to the old labels the new ones. For example,
we have that
\begin{align*}
    \tau_{map} &= \{2 \rightarrow 1 \} \\
    \psi_{map} &= \{(1, 3) \rightarrow (1, 2)\}
\end{align*}
meaning that all the $\tau_2$'s will be relabeled $\tau_1$ and that
$\psi^1_3$ will be relabeled $\psi^1_2$.

\FloatBarrier

\bibliographystyle{ba}
\bibliography{bib_partexch}

\end{document}